\documentclass[journal,comsoc,twocolumn,10pt,twoside]{IEEEtran}
\usepackage{cite}

\ifCLASSINFOpdf
   \usepackage[pdftex]{graphicx}
\else
   \usepackage[dvips]{graphicx}
\fi
\ifCLASSOPTIONcompsoc
  \usepackage[caption=false,font=normalsize,labelfont=sf,textfont=sf]{subfig}
\else
  \usepackage[caption=false,font=footnotesize,labelformat=simple]{subfig}
\fi

\usepackage{stfloats}
\usepackage{url}

\usepackage{epstopdf}
\usepackage{multirow}
\usepackage{diagbox}
\usepackage{color}
\usepackage{amsmath,amsthm,amsfonts,amssymb}
\usepackage{comment}

\begin{document}
%
\title{Risk-Informed Interference Assessment \\for Shared Spectrum Bands: \\A Wi-Fi/LTE Coexistence Case Study}

\author{Andra M. Voicu, Ljiljana Simi\'c, J. Pierre de Vries,\\ Marina Petrova and Petri M\"ah\"onen
\thanks{This paper was presented in part at IEEE DySPAN, Baltimore,  March  2017~\cite{VoicuBaltimore2017}.} 
\thanks{A.~M.~Voicu, L.~Simi\'c, and P. M\"ah\"onen are with the Institute for Networked Systems, RWTH Aachen University, Aachen, Germany (e-mail: avo@inets.rwth-aachen.de; lsi@inets.rwth-aachen.de; pma@inets.rwth-aachen.de). J. P. de Vries is with the Silicon Flatirons Center, University of Colorado, Boulder, U.S. (e-mail: pierredv@hotmail.com). M. Petrova is with the School of Information and Communication Technology, KTH Royal Institute of Technology, 100 44 Stockholm, Sweden (e-mail: \mbox{petrovam@kth.se}).}
}

\maketitle

\begin{abstract}

Interference evaluation is crucial when deciding whether and how wireless technologies should operate.     
In this paper we demonstrate the benefit of risk-informed interference assessment to aid spectrum regulators in making decisions, and to readily convey engineering insight.
Our contributions are: we apply, for the first time, risk assessment to a problem of \mbox{inter-technology} spectrum sharing, i.e. \mbox{Wi-Fi/LTE} in the 5~GHz unlicensed band, and we demonstrate that this method comprehensively quantifies the interference impact.
We perform simulations with our newly publicly-available tool and we consider throughput degradation and fairness metrics to assess the risk for different network densities, numbers of channels, and deployment scenarios. 
Our results show that no regulatory intervention is needed to ensure harmonious technical \mbox{Wi-Fi/LTE} coexistence: for the typically large number of channels available in the 5~GHz band, the risk for \mbox{Wi-Fi} from LTE is negligible, rendering policy and engineering concerns largely moot. As an engineering insight, Wi-Fi coexists better with itself in dense, but better with LTE, in sparse deployments.   
Also, both main LTE-in-unlicensed variants coexist well with Wi-Fi in general. For LTE intra-technology inter-operator coexistence, both variants typically coexist well in the 5 GHz band, but for dense deployments, implementing listen-before-talk causes less interference.    

\end{abstract}

\begin{IEEEkeywords}
coexistence, interference, LTE, risk assessment, spectrum regulation, Wi-Fi.
\end{IEEEkeywords}

%
\IEEEpeerreviewmaketitle

\section{Introduction}

Inter-technology spectrum sharing may generate coexistence problems in bands where mutual interference among different systems occurs. Dynamic spectrum access (DSA) techniques seek to solve such problems by allowing access to the spectrum on a primary-secondary basis, where the primary has priority over secondary systems~\cite{Liang2011}. This problem is managed by each technology individually in the unlicensed bands, where all systems have equal rights to access the spectrum. 

Regardless of the spectrum access rights, inter-technology spectrum sharing raises a \mbox{two-stage} question: (i)~which technologies should/can coexist based on the expected harm of mutual interference, and (ii)~how to manage interactions between technologies on a moment-by-moment basis?    
In this paper we present an extensive case study of applying risk assessment for Wi-Fi/LTE coexistence in the unlicensed bands, in order to evaluate the harm caused by inter-technology interference in shared spectrum bands.

Evaluating coexistence problems due to co- and adjacent channel interference is of interest both to spectrum regulators seeking to establish operational bounds and to engineers designing and managing systems for optimized performance within the regulatory restrictions.
Assessing interference is not a trivial task; consequently, most of the studies manage this complexity by considering \mbox{worst-case} scenarios as the baseline.
Nevertheless, it is not clear how often or under what conditions such worst-case scenarios would occur in practice. Making regulatory decisions based on \mbox{worst-case} analysis may even lead to a complete exclusion of new entrant technologies, so that the second question of interference management becomes irrelevant. As such, comprehensive interference assessment methods are essential for creating a regulatory environment that would enable the deployment of advanced spectrum-sharing techniques, e.g. for \mbox{DSA-like} scenarios. 
Effective interference assessment methods are equally important for the engineers who design, deploy, and manage networks of different technologies coexisting in shared bands, e.g. IEEE 802.11g and n, and \mbox{Wi-Fi/LTE} in the unlicensed bands. Coexistence performance optimization of such networks cannot be conducted under worst-case conditions only.      

In this paper we demonstrate the benefit of risk assessment as a complement to \mbox{worst-case} interference analysis.   
Importantly, risk assessment is a very new method in the fields of communications engineering and spectrum regulation, although it has been used successfully in other fields~\cite{deVries2017}.
We apply risk assessment to a \mbox{Wi-Fi/LTE} coexistence study in the 5~GHz unlicensed band for different network densities, number of channels, and scenarios, both from the point of view of Wi-Fi incumbents and LTE-in-unlicensed entrants.
Our contributions are: (i)~we are the first to apply risk-informed interference assessment to a real-life, topical problem (dealing with inter-technology spectrum sharing with wide relevance for regulatory DSA-like scenarios); and (ii) we demonstrate the benefit of risk assessment as a method that comprehensively and quantitatively characterizes the harm caused by interference in an intuitive and illustrative manner, from both policy and engineering perspectives.
Furthermore, we provide a publicly-available network simulation tool~\cite{iNETS2017} for risk-informed interference assessment of Wi-Fi/LTE coexistence, implementing our simulation model in Section~\ref{section_model}.   

Our analysis shows that no regulatory intervention is needed to ensure harmonious technical coexistence\footnote{Considering economic and policy coexistence issues, e.g. deploying LTE-in-unlicensed for anti-competitive practices, is out of the scope of this paper.}  between \mbox{Wi-Fi/LTE} in the unlicensed bands.
From an engineering perspective, we show that Wi-Fi coexists better with itself and worse with LTE in locally dense deployments, but that the opposite holds in sparse deployments, due to the specifics of \mbox{Wi-Fi's} MAC.
Also, given the large number of available channels expected in practice in the 5~GHz band, there is typically no risk of interference caused by LTE-in-unlicensed entrants, which renders both policy and engineering coexistence issues largely irrelevant. 
In general, both main proposed \mbox{LTE-in-unlicensed} entrant variants coexist equally well with Wi-Fi.  
For LTE intra-technology inter-operator coexistence, both variants typically coexist well in the 5 GHz band, but for very dense deployments, the variant implementing listen-before-talk (LBT) causes less mutual interference between operators.

The remainder of this paper is organised as follows. Section~\ref{section_lte} gives a brief overview of LTE-in-unlicensed and prior work on its coexistence with \mbox{Wi-Fi}. Section~\ref{section_risk} presents the risk-informed interference assessment method. Section~\ref{section_model} presents the simulation and throughput model. Section~\ref{section_results} illustrates and discusses the benefit of applying the risk assessment method for our \mbox{Wi-Fi/LTE} case study, from the point of view of Wi-Fi incumbents. 
Section~\ref{sec_discussion} presents and discusses risk analysis results from the perspective of LTE-in-unlicensed and Section~\ref{section_conclusions} concludes the paper.

\section{LTE-in-unlicensed: The Story so Far}
\label{section_lte}

LTE operation in the unlicensed 5~GHz band has recently been proposed by industry~\cite{Forum2015, Qualcomm2015}. Initially, the unlicensed band is aggregated only for user data transmissions, while the control traffic is sent over the licensed bands for reliability reasons~\cite{3GPP2015}. Two main \mbox{LTE-in-unlicensed} variants with fundamentally different MAC mechanisms have emerged: (i)~\mbox{LTE-U} proposed by the \mbox{LTE-U} Forum~\cite{Forum2015}; and (ii)~Licensed Assisted Access (LAA) first standardized by 3GPP in Release~13~\cite{3GPP2015}. 

\mbox{LTE-U} is based on an adaptive duty cycle MAC mechanism, which adjusts the periodic transmission duration of the devices according to the number of other devices operating in the same channel, such that all devices have an equal share of the channel in time. However, \mbox{LTE-U} devices do not sense and defer to ongoing transmissions before starting their own transmissions, so collisions are likely. \mbox{LTE-U} is a \mbox{pre-standard} version intended for  markets where LBT is not required by regulators (e.g. the U.S.).

LAA is based on LBT, a MAC mechanism in which devices start transmitting only after detecting that the channel is unoccupied. LBT is required by spectrum regulators in some regions (e.g. Europe), so LAA was proposed as a globally applicable standard.  

As \mbox{Wi-Fi} is currently the dominant technology in the 5~GHz band, it has been claimed by some parties (e.g.~\cite{NationalCable&TelecommunicationsAssociation2015}) that introducing \mbox{LTE-in-unlicensed} would harm \mbox{Wi-Fi} operation. On the other hand, proponents have argued that \mbox{LTE-in-unlicensed} would actually improve \mbox{Wi-Fi} performance compared to \mbox{Wi-Fi} coexisting with itself~\cite{Forum2015, Qualcomm2015}. The debate between the two camps led the FCC to issue a Public Notice requesting comments on LTE coexistence in the unlicensed bands~\cite{U.S.FederalCommunicationsCommission2015}, implicitly raising the question of whether regulatory intervention is required to ensure harmonious technical coexistence between \mbox{LTE-in-unlicensed} and \mbox{Wi-Fi}.     

Most existing Wi-Fi/LTE coexistence analyses are not thorough enough to answer the public policy question of whether LTE is friend or foe to \mbox{Wi-Fi} in the unlicensed band. Some existing work lacks a detailed description of algorithms and models (e.g.~\cite{Forum2015}), so that it is difficult to draw generalizable conclusions from the presented results. Other work considers only one main LTE-in-unlicensed variant (\emph{cf}. classification of related work in~\cite{Voicu2016}), so that the results only partially characterize the \mbox{Wi-Fi/LTE} coexistence problem. In our previous work~\cite{Simic2016} we presented the results of a transparent, systematic, and extensive coexistence study and we showed that \mbox{LTE-in-unlicensed} is neither friend nor foe to \mbox{Wi-Fi}. 

In this paper we extend our previous work by conducting, for the first time, a risk assessment of the Wi-Fi/LTE coexistence problem, in order to show the effectiveness of this method for deriving regulatory and engineering insight from quantitative results in a comprehensive, illustrative, and intuitive manner. Furthermore, we extend our throughput model from~\cite{Voicu2016} by incorporating adjacent channel interference and we consider throughput fairness as an additional coexistence performance metric. Finally, we present more detailed results than in~\cite{Voicu2016, Simic2016} by showing the full distributions of our considered metrics.


\begin{table*}[t!]
\caption{Scenarios and entrant variants}
\label{table_1}
\centering
\begin{tabular}{|p{1.5cm}|p{1.5cm}|c|c|}
\hline
	\multicolumn{2}{|c|}{\diagbox[width=4.3cm, height=1cm]{\textbf{PARAMETER}}{\textbf{SCENARIO}}} 
	& \begin{tabular}{c} \emph{\textbf{Indoor/indoor}}\\ \textbf{(indoor incumbent,} \\ \textbf{indoor entrant)}\end{tabular}
	& \begin{tabular}{c} \emph{\textbf{Outdoor/outdoor}}\\ \textbf{(outdoor incumbent,} \\ \textbf{outdoor entrant)}\end{tabular} \\
\hline
	\multicolumn{2}{|c|}{\textbf{Network size}}
	& \begin{tabular}{p{1.2cm}p{4cm}} \textbf{incumbent}:&10~APs\\ \textbf{entrant}:&1--30~APs\end{tabular}	
	& \begin{tabular}{p{1.2cm}p{4cm}} \textbf{incumbent}:&10~APs\\ \textbf{entrant}:&1--10~APs\end{tabular}\\
\hline
	\multicolumn{2}{|c|}{\begin{tabular}{c} \textbf{Maximum number of}\\ \textbf{available channels} \textbf{(Europe)} \end{tabular}}
	& 19
	& 11\\
\hline
	\multirow{10}{*}{\begin{minipage}{2cm}\emph{\textbf{Coexistence}}\\ \emph{\textbf{mechanism}}\end{minipage}} 
	& \begin{minipage}{2cm}\textbf{Channel \\selection}\end{minipage}
	& \multicolumn{2}{l|}{\begin{tabular}{l} \textbf{incumbent}: random \emph{or} single channel\\
	                     \textbf{entrant}: random \emph{or} sense (select channel with fewest incumbent APs) \emph{or} single channel
	                      \end{tabular}}\\
\cline{2-4}                      
	& \textbf{MAC} & \multicolumn{2}{l|}{\begin{tabular}{p{1cm}rl} 
	                                     \textbf{incumbent}: & \textbf{Wi-Fi}: & LBT, CS threshold of -82~dBm for co-channel Wi-Fi devices, \\
	                                                         & & and -62~dBm for co-channel non-Wi-Fi and all adjacent channel devices\\
	                                     \textbf{entrant}: & & \\
	                                     & \textbf{LAA}: & LBT, CS threshold of -62~dBm\\
	                                     & \textbf{LTE-U}: &ON/OFF with adaptive duty cycle based on number of entrant\\
	                                     & & \& incumbent APs within CS range (CS threshold = -62~dBm)\\
	                                     & \textbf{Wi-Fi}: &LBT, CS threshold of -82~dBm for co-channel Wi-Fi devices, \\
	                                                         & & and -62~dBm for co-channel non-Wi-Fi and all adjacent channel devices\\
	                                    \end{tabular}}\\
\hline
	\multicolumn{2}{|c|}{\textbf{PHY}} &  \multicolumn{2}{l|}{\begin{tabular}{p{1cm}rl} 
	                                     \textbf{incumbent}: & \textbf{Wi-Fi}: & IEEE 802.11n spectral efficiency $\rho_{WiFi}$, noise figure NF=15~dB\\
	                                     \textbf{entrant}: & & \\
	                                     & \textbf{LAA}: & LTE spectral efficiency $\rho_{LTE}$, NF=9~dB\\
	                                     & \textbf{LTE-U}: & LTE spectral efficiency $\rho_{LTE}$, NF=9~dB\\
	                                     & \textbf{Wi-Fi}: & IEEE 802.11n spectral efficiency $\rho_{WiFi}$, NF=15~dB 
	                                    \end{tabular}}\\   
\hline      
    \multicolumn{2}{|c|}{\begin{tabular}{c} \textbf{LBT parameters}\\ \textbf{\& assumptions} \end{tabular}} 
           & \multicolumn{2}{l|}{\begin{tabular}{l} binary exponential random backoff with $CW_{min}$=15, $CW_{max}$=1023,\\
                                    time slot duration $\sigma$=9~$\mu$s, SIFS=16~$\mu$s, DIFS=SIFS+2$\sigma$=34~$\mu$s (\emph{cf}. IEEE 802.11) 		   											\end{tabular}}  \\
\hline
    \multicolumn{2}{|c|}{\begin{tabular}{c} \textbf{LBT frame}\\ \textbf{duration $T_f$} \end{tabular}}                                                           
    & \multicolumn{2}{l|}{\begin{tabular}{rl} \emph{\textbf{Wi-Fi}}: & $T_f=fn(rate,~MSDU,~PHY_{header},~MAC_{header})$, \\
                                                                      & $MSDU$=1500~Bytes, $PHY_{header}$=40~$\mu$s, $MAC_{header}$=320~bits (\emph{cf}. IEEE 802.11)\\
                                                \emph{\textbf{LAA}}:  & $T_f$=1~ms (i.e. duration of LTE subframe) \end{tabular}}\\
\hline
     \multicolumn{2}{|c|}{\textbf{Duty cycle ON-time}} 
     &  \multicolumn{2}{l|}{\emph{\textbf{LTE-U}}: 100~ms (i.e. maximum ON-time specified in~\cite{Qualcomm2015})}\\                                 
\hline
     \multicolumn{2}{|c|}{\textbf{User distribution}}
     &  \multicolumn{2}{l|}{1 user per AP}\\
\hline
     \multicolumn{2}{|c|}{\textbf{Traffic model}}
     &  \multicolumn{2}{l|}{downlink full-buffered}\\  
\hline
     \multicolumn{2}{|c|}{\textbf{Channel bandwidth}}
     &  \multicolumn{2}{l|}{20~MHz}\\ 
\hline
     \multicolumn{2}{|c|}{\textbf{Frequency band}}
     &  \multicolumn{2}{l|}{5~GHz (5150--5350 and 5470--5725 MHz)}\\
\hline
	\multicolumn{2}{|c|}{\textbf{AP transmit power}}	
    & \multicolumn{2}{l|}{23~dBm}\\ 
\hline      
\end{tabular}
\end{table*}

\section{Risk-Informed Interference Assessment}
\label{section_risk}
\subsection{Introduction to Risk Assessment}
Risk-informed interference assessment was introduced as a comprehensive, quantitative tool for a spectrum regulator seeking to balance the interests of incumbents, new entrants and the public when deciding whether and how to allocate new radio services~\cite{FCCTAC2015}. It facilitates a balanced assessment of the adverse technical impact of new entrants on incumbents.

Engineering risk assessment, a well-established method used in many industries (e.g. nuclear energy, environmental protection, food safety, etc.~\cite{deVries2017}), considers the likelihood-consequence combinations for multiple hazard scenarios, and complements a ``worst case'' analysis that considers the single scenario with the most severe consequence, regardless of its likelihood. Charts that plot the severity of hazards against their likelihoods are frequently used to visualize and compare the risk of different hazards; see Fig.~\ref{fig_1b}.

To date, quantitative risk assessment has not been used in spectrum management. The author in~\cite{deVries2015} proposed a four-step method for performing risk-informed interference assessment: \textbf{(1)} make an inventory of all significant harmful interference hazard modes; \textbf{(2)} define a consequence metric to characterize the severity of hazards; \textbf{(3)} assess the likelihood and consequence of each hazard mode; and \textbf{(4)} aggregate them into a basis for decision making. In~\cite{deVries2015, Vries2017} it was shown how this method could be used to analyse the risk of cellular interference to weather satellite earth stations for a hypothetical general case. By contrast, we are the first to apply risk-informed interference assessment to a \mbox{real-life} problem and to \mbox{inter-technology} coexistence in the same spectrum band.

\subsection{Applying Risk Assessment to Wi-Fi/LTE Coexistence}

In this paper we evaluate co- and adjacent channel interference among LTE-in-unlicensed entrants and Wi-Fi incumbents by applying risk assessment. In Section~\ref{section_model} we present the interference hazards corresponding to Step~(1). In Section~\ref{metrics_risk} we define the throughput consequence metrics to characterize hazard severity for Step~(2). In Section~\ref{section_results} we demonstrate Steps~(3) and (4) by assessing the hazard modes and by showing the effectiveness of risk assessment when making decisions of regulatory and engineering concern, from the point of view of the Wi-Fi incumbents. In Section~\ref{sec_discussion} we extend our demonstration for Steps~(3) and (4) by presenting the LTE-in-unlicensed perspective.

\subsection{Consequence Metrics for Risk Assessment}
\label{metrics_risk}

In this section we define the consequence metrics to characterize the severity of the interference hazards. 
In the context of Wi-Fi/LTE coexistence we select two throughput metrics\footnote{We note that throughput has been the baseline network performance evaluation metric in general and is also considered the only or primary performance metric in important Wi-Fi/LTE coexistence studies, e.g.~\cite{Forum2015}.
Although delay can be considered a relevant evaluation metric in some cases, it is typically applied for VoIP traffic~\cite{Alliance2016a}, which does not represent the majority of the traffic.} that represent the hazard consequence for the incumbents: (i)~the throughput degradation, which we consider the most relevant metric to quantify whether Wi-Fi gets a fair share of the channel and whether it experiences excessive interference when coexisting with LTE-in-unlicensed, and thus to answer the technical public policy coexistence question; and (ii)~the throughput unfairness among incumbents, which gives insight into engineering optimization of inter-technology coexistence within the given regulatory context. 

We define the throughput degradation of the incumbent access points (APs) when coexisting with entrant APs with respect to two different baselines: (i)~the standalone \mbox{Wi-Fi} incumbent network, in order to capture the general throughput degradation due to network densification; and (ii)~the Wi-Fi incumbent network coexisting with a Wi-Fi entrant network, in order to directly focus on  the question of whether LTE is a better neighbour to Wi-Fi than Wi-Fi is to itself. For an incumbent AP $x$ we estimate the throughput degradation as
\begin{equation}
\label{eq_1}
\Delta R_x=\frac{R_{x,baseline}-R_x}{R_{x,baseline}},
\end{equation} 
where $R_{x,baseline}$ is the baseline throughput of $x$ and $R_x$ is the throughput of $x$ when coexisting with a given entrant variant. 

In order to quantify the throughput fairness among incumbent APs, we apply Jain's fairness index~\cite{Jain1984} for a set of incumbent throughput results corresponding to APs in a single network realization, given by 
\begin{equation}
\label{eq_2}
J=\frac{|\sum^{n}_{x=1} R_{x}|^2}{n\sum^{n}_{x=1}R^2_x},
\end{equation}
where $n$ is the number of incumbents in the network. 

For consistency with data representation in a risk assessment chart (explained in Section~\ref{section_read_chart}), we define the incumbent throughput unfairness as the consequence metric, given by
\begin{equation}
\label{eq_3}
U=1-J.
\end{equation}

When considering the LTE-in-unlicensed perspective in Section~\ref{sec_discussion}, the throughput consequence metrics are analogous to those defined for the incumbents in~\eqref{eq_1}--\eqref{eq_3}.

\section{Simulation \& Throughput Models}
\label{section_model}

\subsection{Simulation Model}
\label{simulation_model}

We assume a population of \mbox{Wi-Fi} incumbent APs coexisting with Wi-Fi, LAA or LTE-U entrant APs in two main scenarios, for realistic network densities, as summarized in Table~\ref{table_1}. 

The incumbent APs and their associated users are always \mbox{Wi-Fi} devices implementing the IEEE 802.11n PHY layer and LBT\footnote{We note that CSMA/CA is a specific variant of the more general LBT mechanism, so we refer to it as LBT. In this paper we assume LBT with binary exponential random backoff throughout.} at the MAC layer with a carrier sense~(CS) threshold of -82~dBm for deferring to co-channel \mbox{Wi-Fi} devices, and -62~dBm for adjacent channel \mbox{Wi-Fi} devices and co- and adjacent channel non-Wi-Fi devices. The entrants are either (i)~LAA implementing the LTE PHY and the LBT MAC mechanism with -62~dBm CS threshold for deferring to all other devices, or (ii)~LTE-U that adapts its duty cycle according to the number of detected APs based on the -62~dBm CS threshold. As the baseline for answering the question of whether LTE-in-unlicensed is friend or foe to Wi-Fi, we also consider (iii)~\mbox{Wi-Fi} entrants.

\begin{figure}[t!]
\centering
\captionsetup[subfigure]{width=0.8\linewidth}
\subfloat[\emph{Indoor/indoor} scenario: the incumbents and entrants are located inside a single-floor building with 20 apartments
(each of 10~m~$\times$~10~m~$\times$~3~m). Each AP and its associated user are randomly placed in a single apartment with up to two AP-user pairs. This figure shows an example of the most dense deployment.]
           {\includegraphics[width=0.9\linewidth]{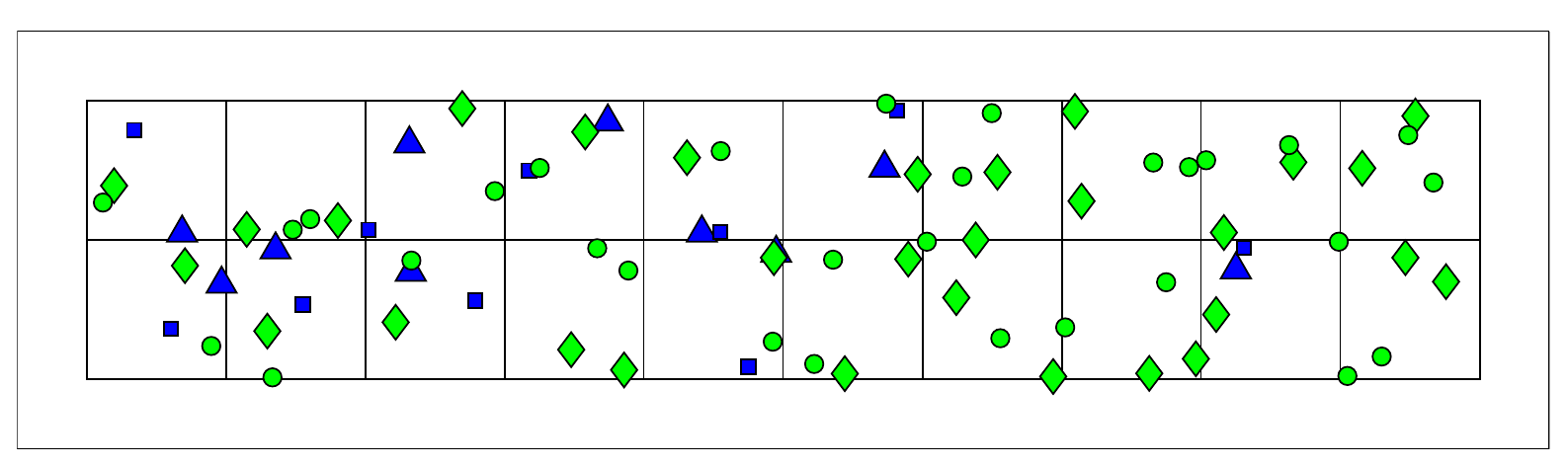} \label{fig_1a}}
\\
\subfloat[\emph{Outdoor/outdoor} scenario: the incumbent and entrant APs are randomly allocated one real outdoor location and are placed at the roof-top level. The outdoor
users are located in the coverage area of and at a maximum distance of 50~m from the AP that they are associated with, at a height of 1.5~m. The length of the buildings is randomly selected between 3--10 apartments and the height is randomly selected between 3--5 floors. The size of the total study area is 346~m~$\times$~389~m, corresponding to the area in London where the real locations of the outdoor APs were observed.]
          {\includegraphics[width=0.8\linewidth]{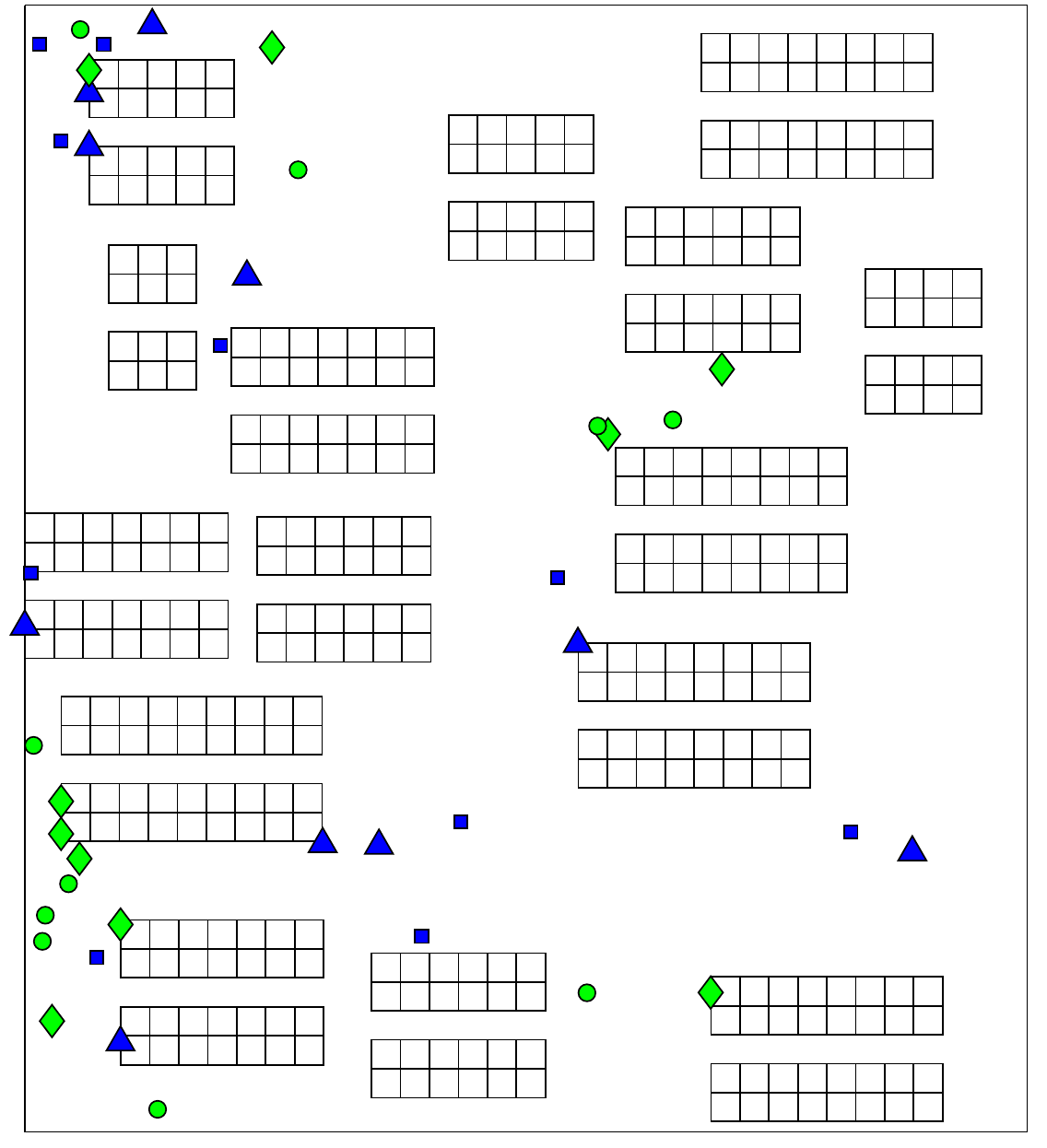} \label{fig_1c}}
\caption{Example network layout based on the 3GPP dual stripe model for indoor deployments and real outdoor
picocell locations for outdoor deployments, for the (a)~\emph{indoor/indoor}, and (b)~\emph{outdoor/outdoor} scenarios, showing locations of incumbent APs ($\color{blue} \blacktriangle$), incumbent
users ($\color{blue} \blacksquare$), entrant APs ($\color{green} \blacklozenge$), and entrant users ($\color{green} \bullet$).}
\label{fig_layout}
\end{figure}

Two main scenarios are considered, where each AP has one associated user, i.e. the \emph{indoor/indoor} scenario where all incumbent and entrant devices are located indoors and the \emph{outdoor/outdoor} scenario where all devices are located outdoors, as in Fig.~\ref{fig_layout}. For the \emph{indoor/indoor} scenario we assume a single-floor building, according to the 3GPP dual stripe model~\cite{Alcatel-Lucent2009}. Each incumbent AP and its associated user are located randomly within a single apartment. The entrant APs and their associated users are first randomly located in unoccupied apartments and then randomly occupy apartments with only one other AP, until all apartments contain up to two APs. This results in network densities of \mbox{600--12000}~APs/km\textsuperscript{2}, as (and more) dense as that seen in contemporary 2.4~GHz deployments, but not yet in 5~GHz~\cite{Achtzehn2013}. For the \emph{outdoor/outdoor} scenario we assume 20 real outdoor base station locations from central London~\cite{MozillaLocationService2015} and we randomly overlay buildings over the area where the real outdoor locations were observed, resulting in network densities of \mbox{7--150}~APs/km\textsuperscript{2}. The associated users are located within the coverage area of the respective APs and at a maximum distance of 50~m. 
As a worst-case interference scenario of low signal attenuation through walls resulting in high interference among APs, we also consider the \emph{indoor/indoor} scenario \emph{without internal walls}.

\begin{table*}[t]
\caption{Parameters for throughput and interference model}
\label{table_equations}
\centering
\begin{tabular}{|c|c|c|}
\hline
	\diagbox[width=2cm, height=1cm]{\textbf{Parameter}}{\textbf{AP type}}
	& \textbf{Incumbent}
	& \textbf{Entrant}\\
\hline
	$S_x$
	& defined in~\cite{Voicu2016}
	& \begin{tabular}{rl} defined in~\cite{Voicu2016}, & if W-Fi/LAA entrant \\
	                     1, & if LTE-U entrant
	                     \end{tabular}\\
\hline	 
    $r_{deg,x}$
    & \begin{tabular}{rl} 0, & if Wi-Fi/LAA entrant \\
                          defined in~\cite{Voicu2016}, & if LTE-U entrant
                          \end{tabular}
    & 0 \\
\hline
    $AirTime_x$
    & \begin{tabular}{r} $\frac{1}{1+|\mathbf{A}_x|+|\mathbf{B}_x|}$, if Wi-Fi/LAA entrant\\
                          $\displaystyle \prod_{y \in \mathbf{B}_x} \Big(1-\frac{1}{1+|\mathbf{C}_y|+|\mathbf{D}_y|}\Big) \times \frac{1}{1+|\mathbf{A}_x|}$, \\
                           if LTE-U entrant
                          \end{tabular}
    & $\frac{1}{1+|\mathbf{A}_x|+|\mathbf{B}_x|}$\\
\hline  
    $\rho_x$
    & $\rho_{WiFi}$~\cite{IEEE2012}
    & \begin{tabular}{rl} $\rho_{WiFi}$, & if Wi-Fi entrant\\
                          $\rho_{LTE}$~\cite{3GPP2009}, & if LAA/LTE-U entrant                           								  \end{tabular}\\
\hline
    $I_u^{co}$
    & $\displaystyle \sum_{z \in (\mathbf{A}^{co} \setminus \mathbf{A}_x^{co}) \cup (\mathbf{B}^{co} \setminus \mathbf{B}_x^{co})} \frac{P_z \times AirTime_z}{L_{u,z}}$
    & \begin{tabular}{rl} $\displaystyle \sum_{z \in (\mathbf{A}^{co} \setminus \mathbf{A}_x^{co}) \cup (\mathbf{B}^{co} \setminus \mathbf{B}_x^{co})} \frac{P_z \times AirTime_z}{L_{u,z}}$, & if Wi-Fi/LAA entrant \\
                          $\displaystyle \sum_{z \in (\mathbf{A}^{co} \setminus \mathbf{A}_x^{co}) \cup (\mathbf{B}^{co})} \frac{P_z \times AirTime_z}{L_{u,z}}$, & if LTE-U entrant
                          \end{tabular}\\
\hline                                                                                                                                                 
    $I_u^{adj}$
    & $\displaystyle \sum_{z \in (\mathbf{A}^{adj} \setminus \mathbf{A}_x^{adj}) \cup (\mathbf{B}^{adj} \setminus \mathbf{B}_x^{adj})} \frac{P_z \times AirTime_z}{L_{u,z} \times ACIR_{u,z}}$
    & \begin{tabular}{rl} $\displaystyle \sum_{z \in (\mathbf{A}^{adj} \setminus \mathbf{A}_x^{adj}) \cup (\mathbf{B}^{adj} \setminus \mathbf{B}_x^{adj})} \frac{P_z \times AirTime_z}{L_{u,z} \times ACIR_{u,z}}$, & if Wi-Fi/LAA entrant \\
                          $\displaystyle \sum_{z \in (\mathbf{A}^{adj} \setminus \mathbf{A}_x^{adj}) \cup (\mathbf{B}^{adj})} \frac{P_z \times AirTime_z}{L_{u,z} \times ACIR_{u,z}}$, & if LTE-U entrant
                          \end{tabular}\\ 
\hline                          
\end{tabular}
\end{table*}

Each incumbent AP randomly selects one of the available channels. The entrants either randomly select a channel, i.e. \emph{random}, or apply \emph{sense}, i.e. they randomly select a channel unoccupied by incumbents. We assume the maximum number of channels in the 5~GHz band in Europe to be typically available in practice (i.e. 19 indoor and 11 outdoor channels), or only the 4 non-DFS channels, corresponding to less likely cases of either legacy devices that do not implement DFS, or devices with faulty DFS implementation (e.g. erroneously detecting radar channels as occupied). As a \mbox{worst-case} of high local AP density corresponding to a high level of interference, we also consider the \emph{single channel} case.

For the indoor links we assume a multi-wall-and-floor model (MWF) model~\cite{LottRhode2001} and for the outdoor links the \mbox{ITU-R} model for line-of-sight (LOS) propagation  within street canyons and for non-line-of-sight (NLOS) with over roof-top propagation~\cite{ITU-R2013}. The model also takes into account log-normal shadowing with a standard deviation of 4~dB for indoor links and 7~dB for outdoor links~\cite{3GPP2010}. 

We perform extensive Monte Carlo simulations in MATLAB with 3000 network realizations for the \emph{indoor/indoor} scenario and \emph{indoor/indoor} scenario \emph{without internal walls}, and 1500 realizations for the \emph{outdoor/outdoor} scenario.
We assume downlink saturated traffic (i.e. most challenging coexistence case) and we evaluate the network performance based on the downlink throughput per AP, estimated at the associated user.\footnote{For multiple users associated to a single AP, the user throughput is obtained by dividing the \mbox{per-AP} throughput to the number of associated users.}   

\subsection{Throughput Model}

Our throughput and interference model for co-channel interference is described in detail in~\cite{Voicu2016} and in this paper we apply it to both co- and adjacent channel interference. 

For \mbox{Wi-Fi} and LAA, we assume the LBT mechanism does not allow co- and adjacent channel APs within CS range\footnote{The CS range within which co- and adjacent channel APs are located is defined according to the respective CS thresholds given in Section~\ref{simulation_model}. We note that the adjacent channel interference ratio (ACIR) is taken into account for adjacent channel power calculations.} of each other to transmit simultaneously. Each of these APs is thus allowed to transmit for only an approximately equal fraction of time. The co- and adjacent channel APs located outside the CS range interfere by decreasing the signal-to-interference-and-noise-ratio (SINR) at the associated user.

For LTE-U, the adaptive duty cycle MAC mechanism adjusts the duty cycle of each AP based on the number of co- and adjacent channel APs detected within the CS range. However, the LTE-U APs within the same CS range may interfere with each other, as they do not check if the channel is unoccupied before transmitting. Instead, they transmit periodically, where we assume uncoordinated LTE-U APs that randomly select the starting moment of their duty cycle period, so that their transmissions may overlap in time. 
The \mbox{Wi-Fi} incumbents sense the medium unoccupied by coexisting LTE-U entrants for a duration determined by the entrants' adaptive duty cycle, and the likelihood of their overlapping transmissions. Consequently, when coexisting with LTE-U entrants, the incumbents detect the medium unoccupied for a different fraction of time than when coexisting with LAA entrants. 
The co- and adjacent channel LTE-U APs located outside the CS range decrease the SINR at the associated incumbent or entrant user. 

In general we estimate the downlink throughput of an AP~$x$ according to our model in~\cite{Voicu2016} as
\begin{equation}
\begin{aligned}
R_x &=S_x \times (1-r_{deg,x}) \times AirTime_x \times \rho_{x}(SINR_u),
\end{aligned}
\end{equation}     
where $S_x$ is the LBT MAC protocol efficiency accounting for sensing time and collisions between LBT frames based on Bianchi's model~\cite{Bianchi2000}, $r_{deg,x}$ is the additional throughput degradation due to collisions between LBT and duty cycle frames, $AirTime_x$ is the fraction of time that AP $x$ is allowed to transmit according to its own and the other \mbox{within-CS-range} APs' MAC mechanisms,  $\rho_{x}$ is the auto-rate function mapping the SINR to the bit rate, and $SINR_u$ is the SINR at the associated user $u$ of $x$. A mathematical description of these parameters in given in Table~\ref{table_equations}, where $\mathbf{A}_x$ is the set of co- and adjacent channel incumbent APs within CS range of $x$, $\mathbf{B}_x$ is the set of co- and adjacent channel entrant APs within CS range of $x$, $|\mathbf{A}_x|$ is the number of co- and adjacent channel incumbent APs within the CS range of $x$, $|\mathbf{B}_x|$ is the number of co- and adjacent channel entrant APs within the CS range of $x$, $|\mathbf{C}_y|$ is the number of co- and adjacent channel incumbent APs within CS range of AP $y$, and $|\mathbf{D}_y|$ is the number of co- and adjacent channel entrant APs within CS range of AP~$y$.  

We estimate the SINR at the associated user $u$ as
\begin{equation}
SINR_u=\frac{P_x(L_{u,x})^{-1}}{I_u^{co} + I_u^{adj} + N_0},
\end{equation}
where $P_x$ is the transmit power of AP $x$, $L_{u,x}$ is the propagation loss between user $u$ and AP $x$, $I_u^{co}$ is the interference from co-channel APs, $I_u^{adj}$ is the interference from adjacent channel APs, and $N_0$ is the background noise (assumed -174~dBm/Hz). A mathematical description of these terms is given in Table~\ref{table_equations}, where $\mathbf{A}^{co}$ is the set of all co-channel incumbent APs of $x$, $\mathbf{A}^{co}_x$ is the set of co-channel incumbent APs within CS range of $x$, $\mathbf{B}^{co}$ is the set of all co-channel entrant APs of $x$, $\mathbf{B}^{co}_x$ is the set of co-channel entrant APs within CS range of $x$, $\mathbf{A}^{adj}$ is the set of all adjacent channel incumbent APs of $x$, $\mathbf{A}^{adj}_x$ is the set of adjacent channel incumbent APs within CS range of $x$, $\mathbf{B}^{adj}$ is the set of all adjacent channel entrant APs of $x$, $\mathbf{B}^{adj}_x$ is the set of adjacent channel entrant APs within CS range of $x$, $P_z$ is the transmit power of AP $z$, $AirTime_z$ is the fraction of time AP $z$ may transmit (defined similarly as $AirTime_x$), $L_{u,z}$ is the propagation loss between $z$ and $u$, and $ACIR_{u,z}$ is the adjacent channel interference ratio given by $z$'s transmitter at $u$'s receiver when operating on adjacent channels. We assume the model in~\cite{3GPP2015} defining $ACIR_{u,z}$ as
\begin{equation}
ACIR_{u,z}=\frac{1}{\frac{1}{ACLR_z} + \frac{1}{ACS_u}},
\end{equation}    
where $ACLR_z$ is the adjacent channel leakage ratio of transmitter $z$ and $ACS_u$ is the adjacent channel selectivity of receiver $u$. For Wi-Fi APs and users we assume $ACLR_z$=26~dB and $ACS_x$=$ACS_u$=22~dB, corresponding to the least efficient Wi-Fi transmitter and receiver,\footnote{We note $ACS_x$ is needed as the power received from adjacent channels is also estimated at AP $x$, since its level may be high enough, such that $x$ may detect the channel busy and may share its channel in time with (or adapt its duty cycle according to) transmissions in the adjacent channels.} whereas for the LTE-in-unlicensed variants we assume $ACLR_z$=45~dB, $ACS_x$=46~dB, and $ACS_u$=22~dB~\cite{3GPP2015}, corresponding to the most efficient LTE AP transmitter and receiver, and the same LTE user receiver as for Wi-Fi.


\section{Risk Assessment from the Wi-Fi Incumbent Perspective}
\label{section_results}

In this section we present a selection of our simulation results that illustrate the effectiveness of risk assessment for Wi-Fi/LTE coexistence. Specifically, we evaluate the risk of co- and adjacent channel interference for the \mbox{Wi-Fi} incumbents 
and we show its relevance for spectrum regulators, i.e. in deciding whether regulatory action is required to ensure harmonious inter-technology coexistence, and for engineers designing and optimizing such networks.

We apply the consequence metrics defined in Section~\ref{metrics_risk} as follows. The throughput degradation is estimated for each incumbent AP in each Monte Carlo network realization, resulting in a distribution of throughput degradation over all incumbents in all network realizations. Jain's unfairness is estimated for each network realization, over the set of incumbent throughput values within a single network realization, resulting in a distribution of unfairness over all network realizations.   

In this section we first discuss how to read risk assessment charts in general and for our case study. Then we focus on individually assessing the risk of interference for different network densities, channel availability, and deployment scenarios, from the Wi-Fi incumbent perspective.  

\subsection{Reading Risk Assessment Charts}
\label{section_read_chart}

Risk assessment representations in general are likelihood-consequence charts where the curves show an increasing risk of harm from the lower left corner to the upper right corner, as indicated by the red arrow in the example of Fig.~\ref{fig_1b}.
For the Wi-Fi/LTE coexistence case, the likelihood-consequence charts illustrate the risk of interference that the incumbents suffer when coexisting with different entrants, by following the general rule of increased risk towards the upper right corner. We represent the likelihood as the CCDF of the throughput consequence metrics, for consistency with this rule.

In our figures showing throughput degradation, e.g. Figs.~\ref{fig_1b} and \ref{fig_1c}, a positive throughput degradation is equivalent to an actual decrease in throughput compared with the considered baseline, whereas a negative throughput degradation shows an increase in throughput. 

\subsection{Effect of Network Density} 
\label{section_netdensit}
 
\begin{figure}[!t]
\centering
\subfloat[Conventional representation]{\includegraphics[width=0.9\columnwidth]{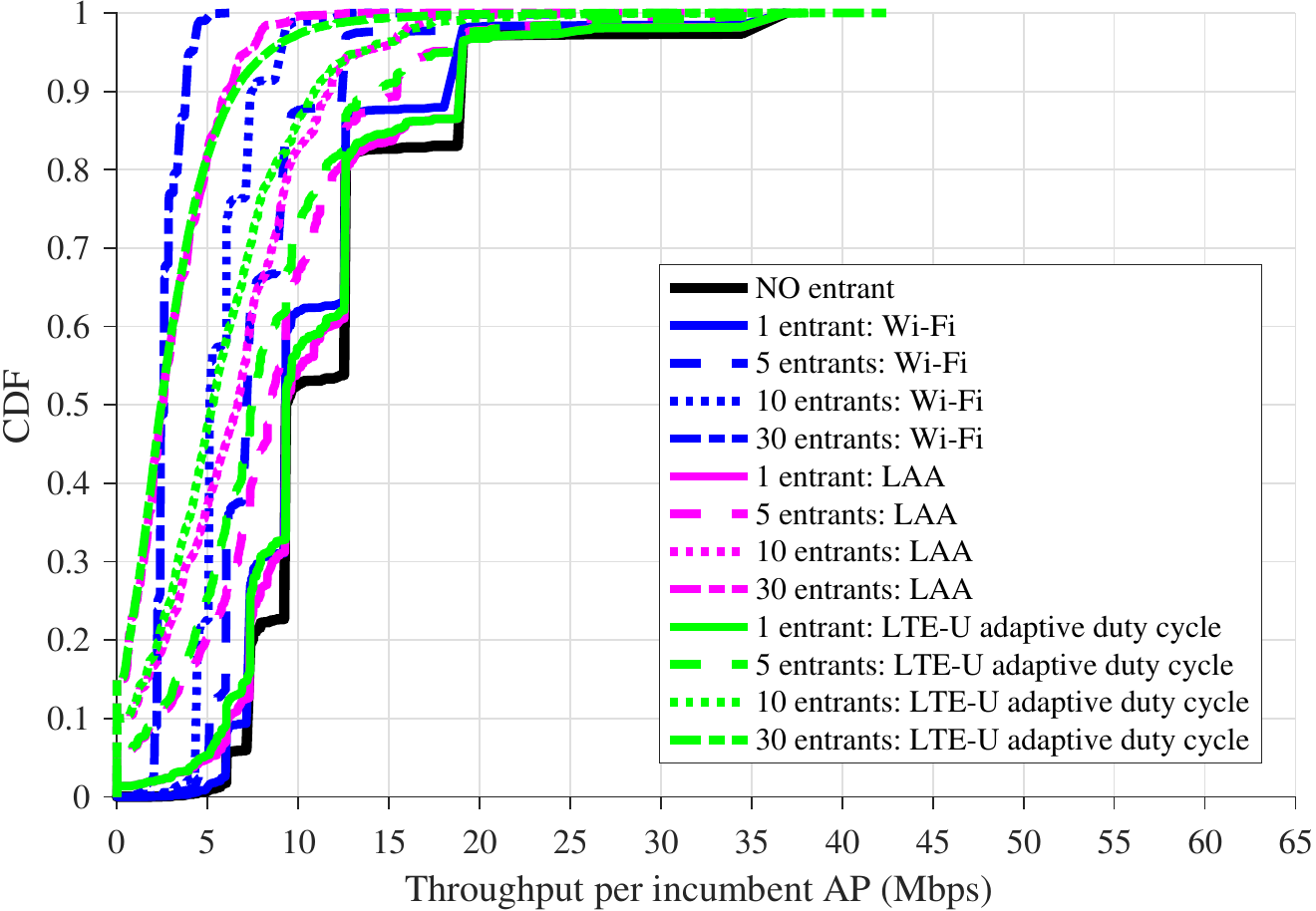} \label{fig_1a}}
\\
\subfloat[Risk representation]{\includegraphics[width=0.9\columnwidth]{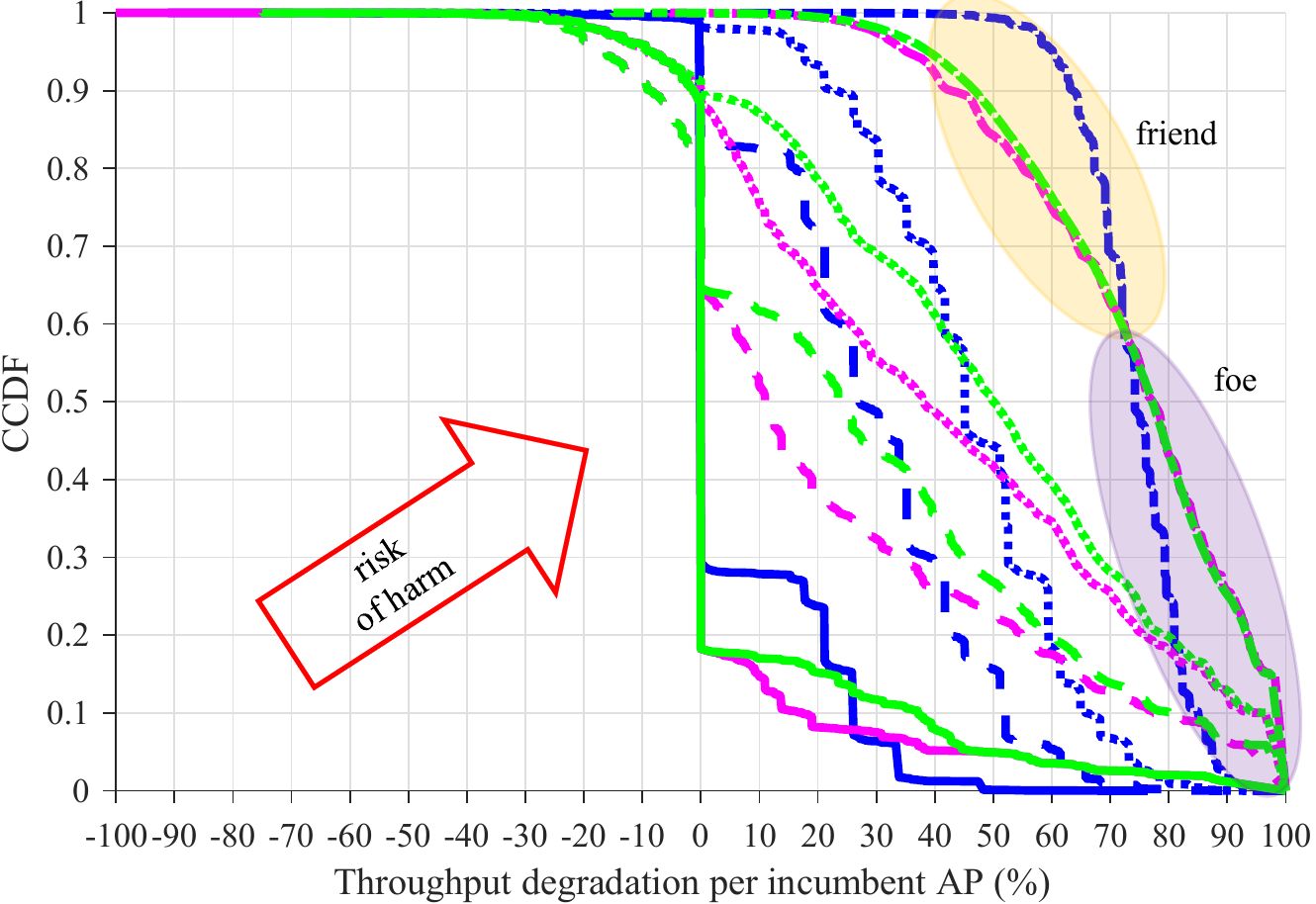} \label{fig_1b}}
\\
\subfloat[Risk representation]{\includegraphics[width=0.9\columnwidth]{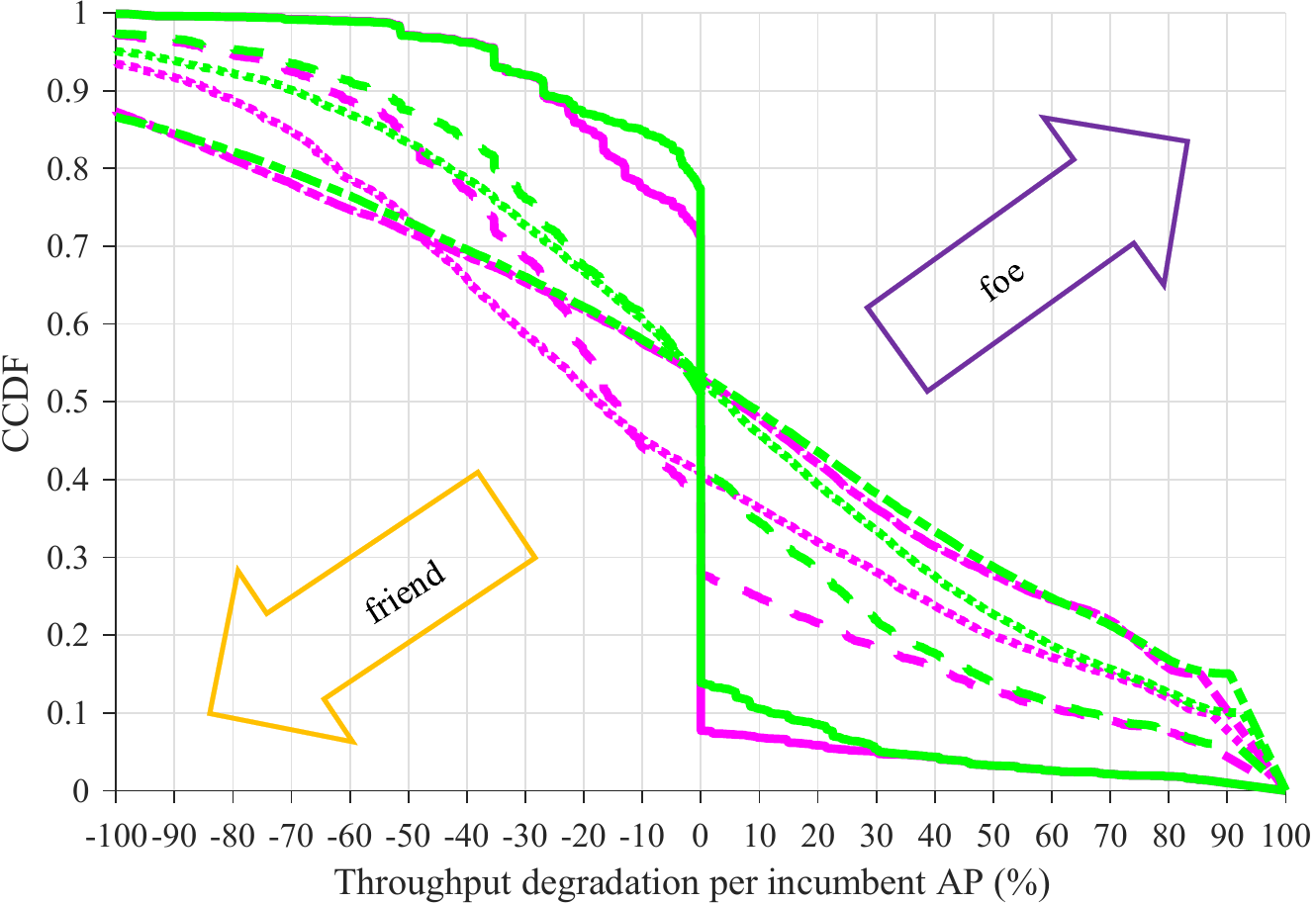} \label{fig_1c}}
\caption{Example of conventional (a) and risk (b) and (c) representations of incumbent AP performance results for the \emph{indoor/indoor} scenario, for \emph{single channel}, for 10 incumbent and 0--30 entrant APs, as (a)~distribution of throughput per incumbent AP; (b)~distribution of throughput degradation per incumbent AP with the standalone incumbents as baseline; and (c)~distribution of throughput degradation per incumbent AP with the incumbents coexisting with \mbox{Wi-Fi} entrants as baseline.}
\label{fig_1}
\end{figure}  
 
In this section we demonstrate the advantage of risk over conventional representation of our coexistence study results when assessing interference for various network densities.  
 
Fig.~\ref{fig_1} shows an example of conventional and risk representations  of incumbent throughput performance results, for the \emph{indoor/indoor} scenario with 10 incumbents and \mbox{0--30 entrants}, for \emph{single channel} (i.e. co-channel interference only).
Specifically, Fig.~\ref{fig_1a} shows an example of a conventional representation as the CDF of the incumbent AP throughput $R_x$. When the number of entrants increases from 0 to 30, the incumbent throughput decreases from e.g. 10 to 2.5~Mbps for the median value. 
Also, Fig.~\ref{fig_1a} shows that for a fixed number of entrants, the throughput of incumbents coexisting with \mbox{Wi-Fi} entrants is sometimes higher and sometimes lower than when coexisting with LAA or LTE-U entrants. This suggests that \mbox{LTE-in-unlicensed} entrants are sometimes friend and sometimes foe to Wi-Fi, but does not readily provide further insight.
Although such a representation of the absolute throughput is important for coexistence cases since it provides the baseline for calculating the throughput degradation as a relative metric, the performance degradation caused by various entrants cannot be quantified in a straightforward way.  

Fig.~\ref{fig_1b} shows the results in Fig.~\ref{fig_1a} in the form of a likelihood-consequence chart, i.e. the CCDF vs. incumbent throughput degradation with the standalone incumbent throughput (i.e. no entrant) as baseline. 
Fig.~\ref{fig_1b} shows in general that the risk increases significantly when the number of entrants increases, irrespective of the entrant technology. The median incumbent throughput degradation increases from 0\% to 75\% for 0 to 30 entrants.
Also, for each number of entrant APs there is a switching point where the order of the curves corresponding to \mbox{Wi-Fi} and LAA or LTE-U is reversed. Consequently, the risk of incumbent throughput degradation when coexisting with \mbox{Wi-Fi} entrants is sometimes higher and sometimes lower than when coexisting with LTE-in-unlicensed. LTE-in-unlicensed is thus neither consistently friend nor foe to \mbox{Wi-Fi}, suggesting the engineering policy question is moot.  

From a more detailed engineering perspective, it is evident from Fig.~\ref{fig_1b} that the Wi-Fi entrants pose greater risk in case of lower negative impact, whereas the LAA and LTE-U entrants pose greater risk in case of higher negative impact. Let us consider the example case of 30 entrants, where the switching point occurs at a throughput degradation of 72\%. For a throughput degradation lower than 72\%, the risk posed by \mbox{Wi-Fi} entrants is higher than for LAA or LTE-U entrants, whereas for a throughput degradation higher than 72\% the opposite holds. This effect occurs due to the value of the CS threshold according to which the incumbent APs defer to the entrants, i.e. the incumbents apply a \mbox{-82}~dBm and \mbox{-62}~dBm threshold to defer to \mbox{Wi-Fi} and LTE-in-unlicensed entrants, respectively (\emph{cf}. IEEE 802.11). 
For the lower CS threshold the incumbents are more conservative and avoid strong interference, by deferring to more entrants and transmitting less often. A lower CS threshold is thus suitable for (locally) dense deployments with strong interference, whereas it causes the incumbents to defer unnecessarily in sparse deployments with low interference. The opposite holds for a higher CS threshold.      

Fig.~\ref{fig_1b} also shows that for a fixed number of entrants, the throughput degradation from \mbox{LTE-U} and LAA is similar, so LTE-U and LAA are almost equally good neighbours to \mbox{Wi-Fi}. The risk is somewhat higher from LTE-U than LAA, due to the additional collisions in term $r_{x,deg}$ and the adjustment of the entrant duty cycle based on the number of devices detected by the entrants only. Consequently, some incumbents are allowed to transmit for a lower fraction of time than their equal share when considering the number of APs in their own CS range.\footnote{The opposite effect was shown in~\cite{Voicu2016} for low incumbent and high entrant densities, where the likelihood of short duty cycles and overlapping entrant transmissions is higher, such that the incumbents find the medium unoccupied by entrants for a longer fraction of time.}

Finally, Fig.~\ref{fig_1b} shows that some of the incumbents have a negative throughput degradation when coexisting with entrants compared with the standalone (i.e. no entrant) network. These cases are due to hidden nodes that are continuous sources of interference in the standalone incumbent network, but that interfere only for a fraction of time when they defer to entrants deployed in the coexistence cases. 
However, the negative throughput degradation is in some cases an artefact of our throughput model, where the MAC efficiency term $S_x$ is averaged over the entire CS range, sometimes resulting in higher average values for the incumbents when LTE-in-unlicensed entrants with higher $S_x$ are located within the CS range. 

In order to focus directly on the question of whether LTE-in-unlicensed is friend or foe to Wi-Fi, Fig.~\ref{fig_1c} shows an alternative risk representation of Fig.~\ref{fig_1b}, where the baseline for incumbent throughput degradation is the incumbent throughput when coexisting with \mbox{Wi-Fi} entrants. A positive throughput degradation thus corresponds to LTE being foe, whereas a negative throughput degradation corresponds to LTE being friend to Wi-Fi in unlicensed bands.
For a given number of entrants, the percentage of incumbents for which the entrants are friends or foes is similar, with up to 50\% being friends and 50\% foes for 30 entrants. This clearly shows that for the typical \emph{indoor/indoor} scenario no regulatory intervention is required. 
In the rest of this paper we focus on the throughput degradation given the standalone incumbent network as baseline (as in Fig.~\ref{fig_1b}), as this case provides better insight into the more general network densification problem.

\begin{figure}[!t]
\centering
\subfloat[Conventional representation]{\includegraphics[width=0.9\columnwidth]{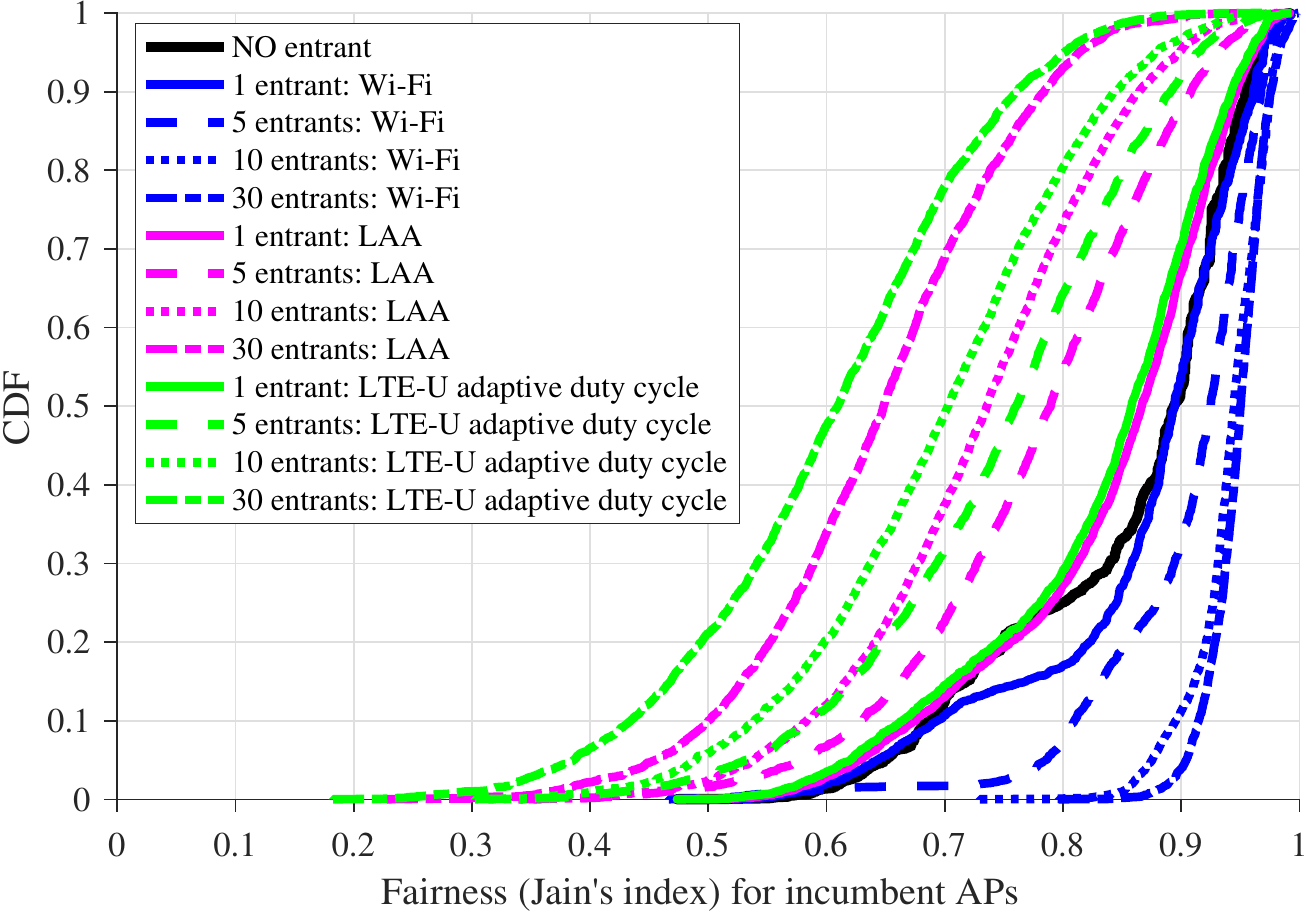} \label{fig_2a}}
\\
\subfloat[Risk representation]{\includegraphics[width=0.9\columnwidth]{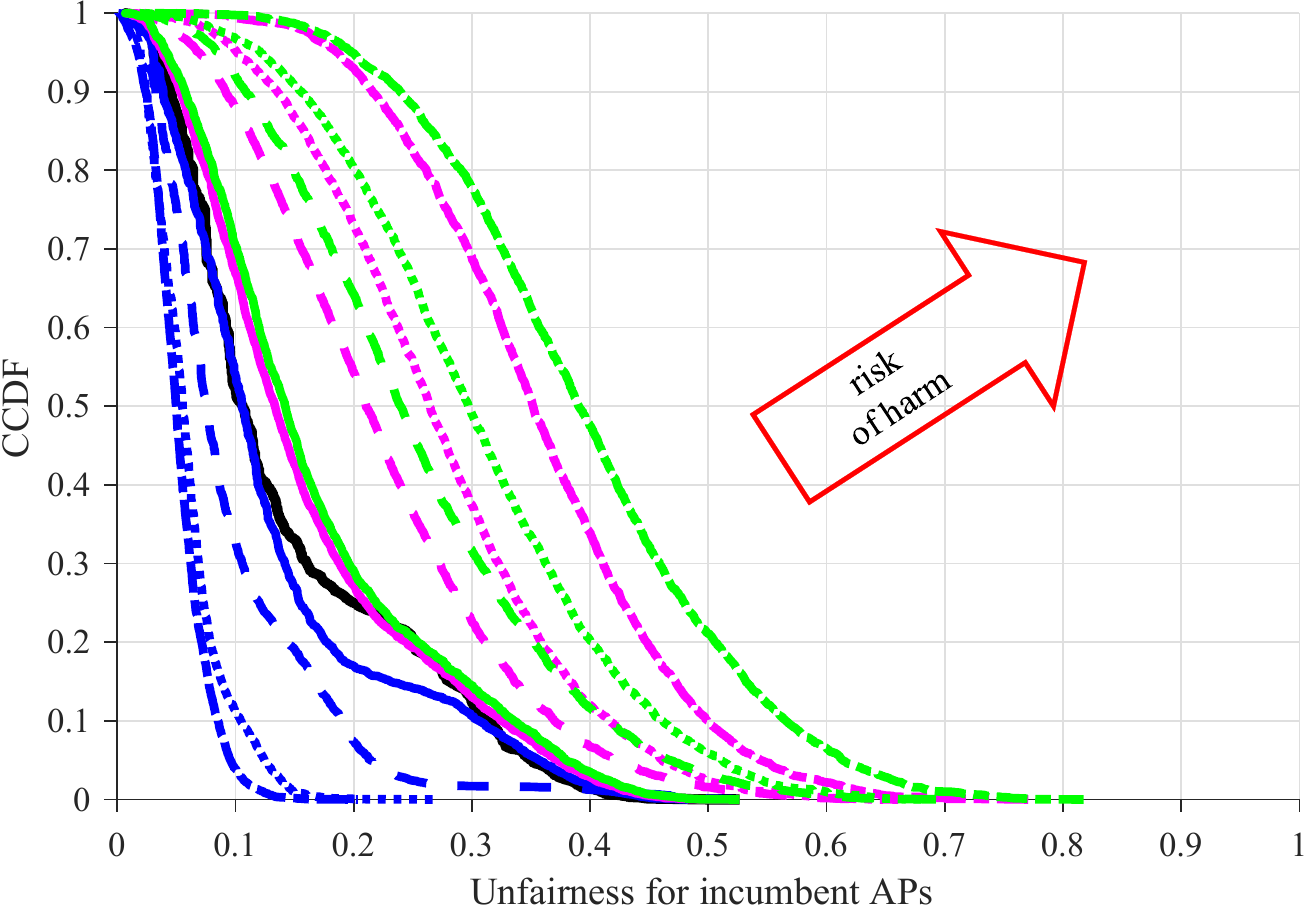} \label{fig_2b}}
\caption{Example of conventional (a) and risk (b) representations of incumbent AP performance results for the \emph{indoor/indoor} scenario, for \emph{single channel}, for 10 incumbent and 0--30 entrant APs, as (a)~distribution of Jain's fairness index for incumbent APs in each network realization; and (b)~distribution of Jain's unfairness index for incumbent APs in each network realization.}
\label{fig_2}
\end{figure}

Let us now consider the second consequence metric, i.e. Jain's unfairness. Fig.~\ref{fig_2} shows an example of conventional and risk representations  of Jain's fairness/unfairness among incumbents, for the \emph{indoor/indoor} scenario with 10 incumbents and 0--30 entrants, corresponding to the throughput degradation in Fig.~\ref{fig_1}. Specifically, Fig.~\ref{fig_2a} shows a conventional representation as the CDF of the fairness index $J$. For consistency with the likelihood-consequence charts, Fig.~\ref{fig_2b} shows the same results as Fig.~\ref{fig_2a} in the form of CCDF of throughput unfairness $U$, where the risk increases towards the upper right corner. We will thus comment only on Fig.~\ref{fig_2b}. For a fixed number of entrants, the risk of incumbent unfairness is higher for LAA or LTE-U entrants than for Wi-Fi entrants, consistent with our results in Fig.~\ref{fig_1b}, which show that the risk of high throughput degradation is higher for LTE-in-unlicensed, resulting in larger variation of the throughput degradation. Also, the risk of unfairness increases with the number of entrants for LAA or LTE-U, whereas it decreases for Wi-Fi, given the different CS thresholds that the incumbents apply. Moreover, the risk of unfairness decreases for Wi-Fi below the risk for the standalone incumbent network. Also, \mbox{LTE-U} has a higher risk of unfairness compared with LAA, consistent with its higher throughput degradation for only some incumbents.

Importantly, our results show that for \emph{single channel} the risk is qualitatively different for the two considered consequence metrics. The risk of throughput degradation (relevant for the engineering policy question) in Fig.~\ref{fig_1b} is sometimes higher and sometimes lower for coexistence with LAA or LTE-U than with \mbox{Wi-Fi} (i.e. LTE-in-unlicensed is sometimes friend and sometimes foe). By contrast, the risk of Jain's unfairness among incumbents (relevant for engineering performance optimization) in Fig.~\ref{fig_2b} is always higher with LAA or LTE-U than with \mbox{Wi-Fi} (i.e. with Wi-Fi, all incumbents are affected in a similar way).
This illustrates the importance of choosing a metric that effectively quantifies policy goals, as different metrics, encoding different values, may lead to different conclusions.

\subsection{Effect of Channel Availability}

\begin{figure}[!t]
\centering
\subfloat[]{\includegraphics[width=0.9\columnwidth]{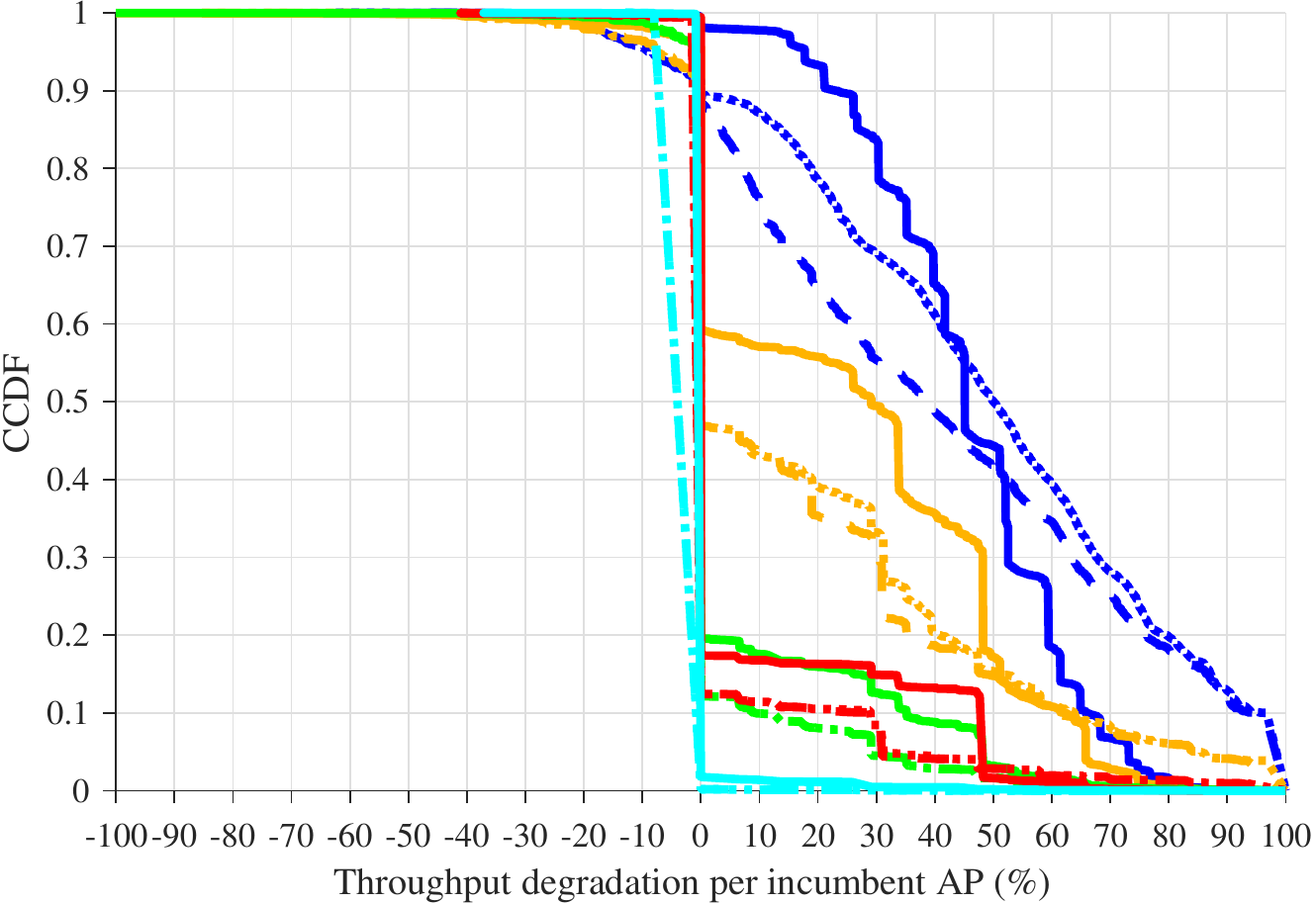} \label{fig_3a}}
\\
\subfloat[]{\includegraphics[width=0.9\columnwidth]{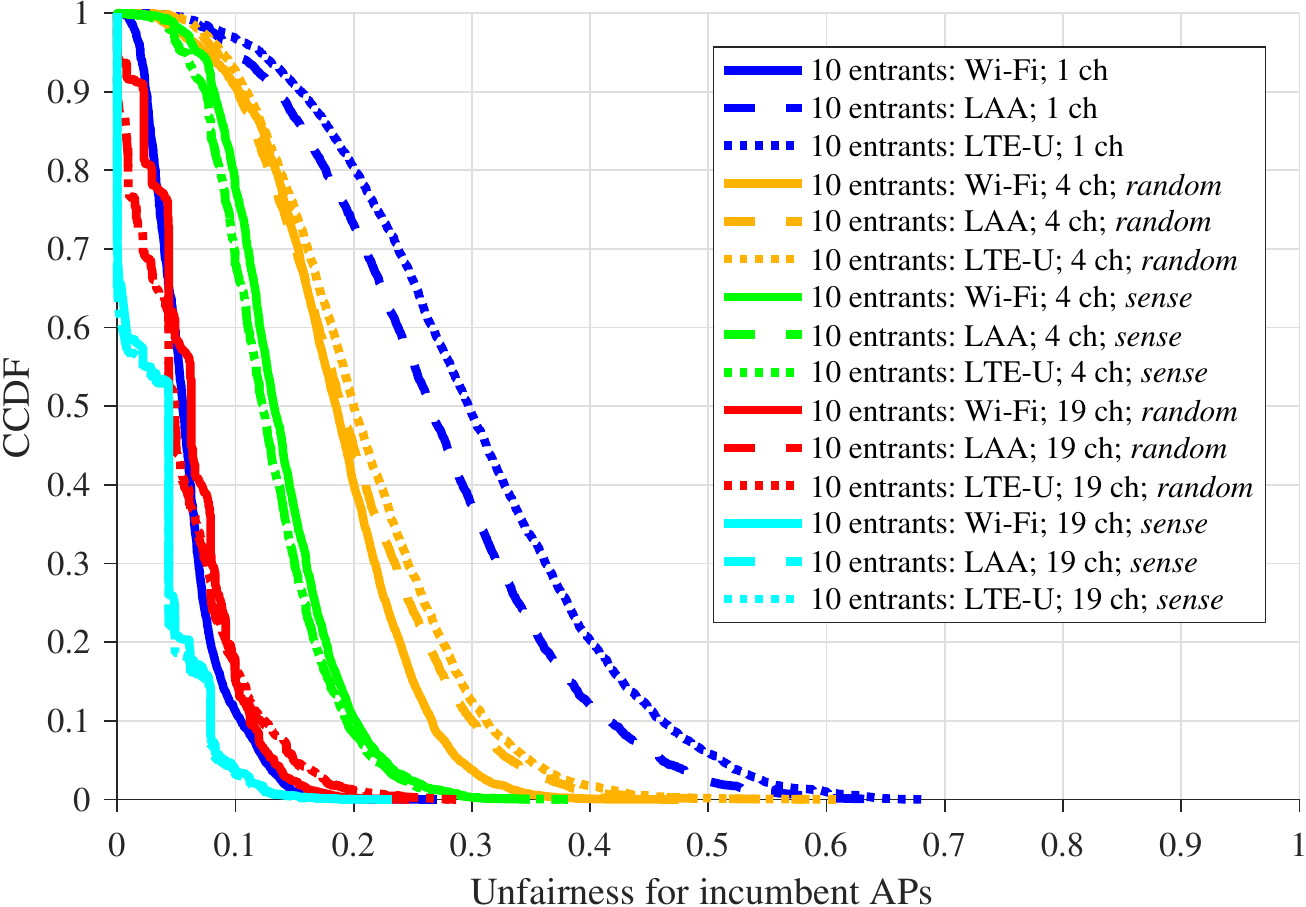} \label{fig_3b}}
\caption{Risk representation of incumbent AP performance results for the \emph{indoor/indoor} scenario, for \textbf{different number of channels}, for 10 incumbent and 10 entrant APs, as (a)~distribution of throughput degradation per incumbent AP with the standalone incumbents as baseline; and (b)~distribution of Jain's unfairness index for incumbent APs in each network realization.}
\label{fig_3}
\end{figure}

In this section we assess the risk of interference for the \mbox{Wi-Fi/LTE} coexistence case, for different numbers of channels (i.e. with co- and adjacent channel interference) and channel selection schemes. 
Fig.~\ref{fig_3} shows the incumbent throughput degradation and Jain's unfairness, for the \emph{indoor/indoor} scenario, for 10 incumbents and 10 entrants (i.e. an example with a single AP in each apartment), and 1, 4 and 19 channels with \emph{sense} and \emph{random}. 
The risk of throughput degradation in Fig.~\ref{fig_3a} increases when the number of channels decreases, from 0\% median throughput degradation for 19 channels to 40-50\% median throughput degradation for \emph{single channel}.

For \emph{sense} with the maximum number of 19 channels (typically available in practice), near-perfect coexistence is ensured between incumbents and entrants (i.e. 0\% incumbent throughput degradation), due to the large number of unoccupied channels that the entrants can select from. Also, \emph{random} with 19 channels has similar performance, with only a small percentage of incumbent APs suffering a rather low throughput degradation. This shows that no regulatory or engineering action is needed to ensure harmonious coexistence. As an engineering insight, Fig.~\ref{fig_3} reveals that \emph{sense} does not bring significant benefit for such a high number of channels. 

For non-DFS devices operating on 4 channels with the entrants implementing \emph{sense}, the throughput degradation is similar to the one for 19 channels, whereas for 4 channels with \emph{random} the throughput degradation increases significantly, showing that engineers should implement \emph{sense} for the rare cases of such a low number of channels. 
Also, the switching point delimiting the friend/foe entrants (explained in Section~\ref{section_netdensit}) is visible for 4 channels \emph{random} and for \emph{single channel}; for the other cases LTE-in-unlicensed is an equally good or better neighbour to Wi-Fi than Wi-Fi is to itself. 

Fig.~\ref{fig_3b} shows the CCDF of Jain's incumbent unfairness for 1 to 19 channels, where the unfairness increases when the number of channels decreases, with the exception of \mbox{Wi-Fi}, for \emph{single channel}. 
The highest unfairness is caused by the LAA or LTE-U entrants for \emph{single channel}, but for 4 and 19 channels the unfairness is similar to the one caused by \mbox{Wi-Fi} entrants, consistent with the similar throughput degradation results for all entrant technologies for these number of channels. Importantly, for the typical 19 and also for 4 non-DFS available channels, both consequence metrics consistently show that there is no coexistence problem relevant for engineering policy or engineering optimization.  

\subsection{Effect of Deployment Scenario} 

\begin{figure}[!t]
\centering
\subfloat[]{\includegraphics[width=0.9\columnwidth]{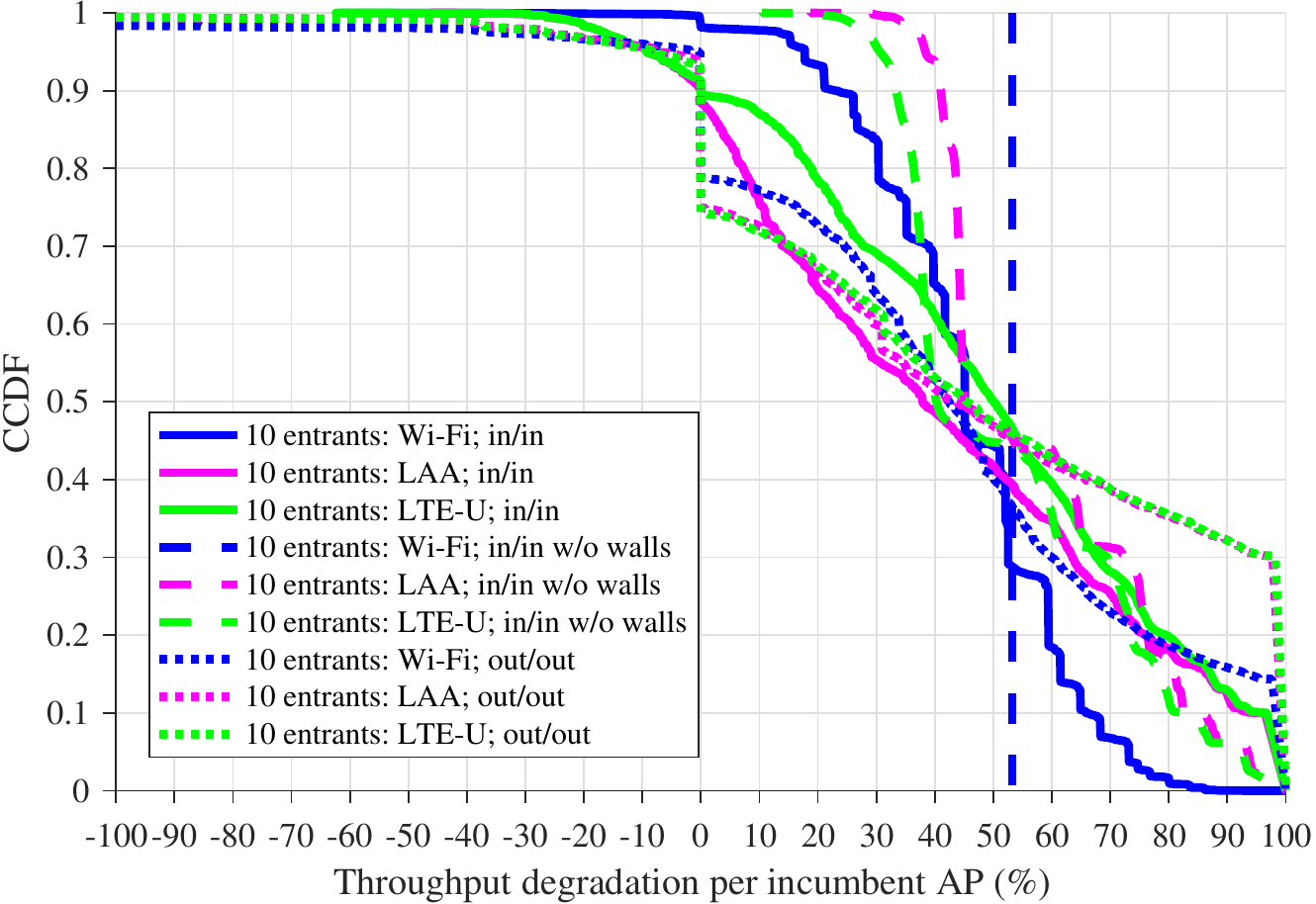} \label{fig_4a}}
\\
\subfloat[]{\includegraphics[width=0.9\columnwidth]{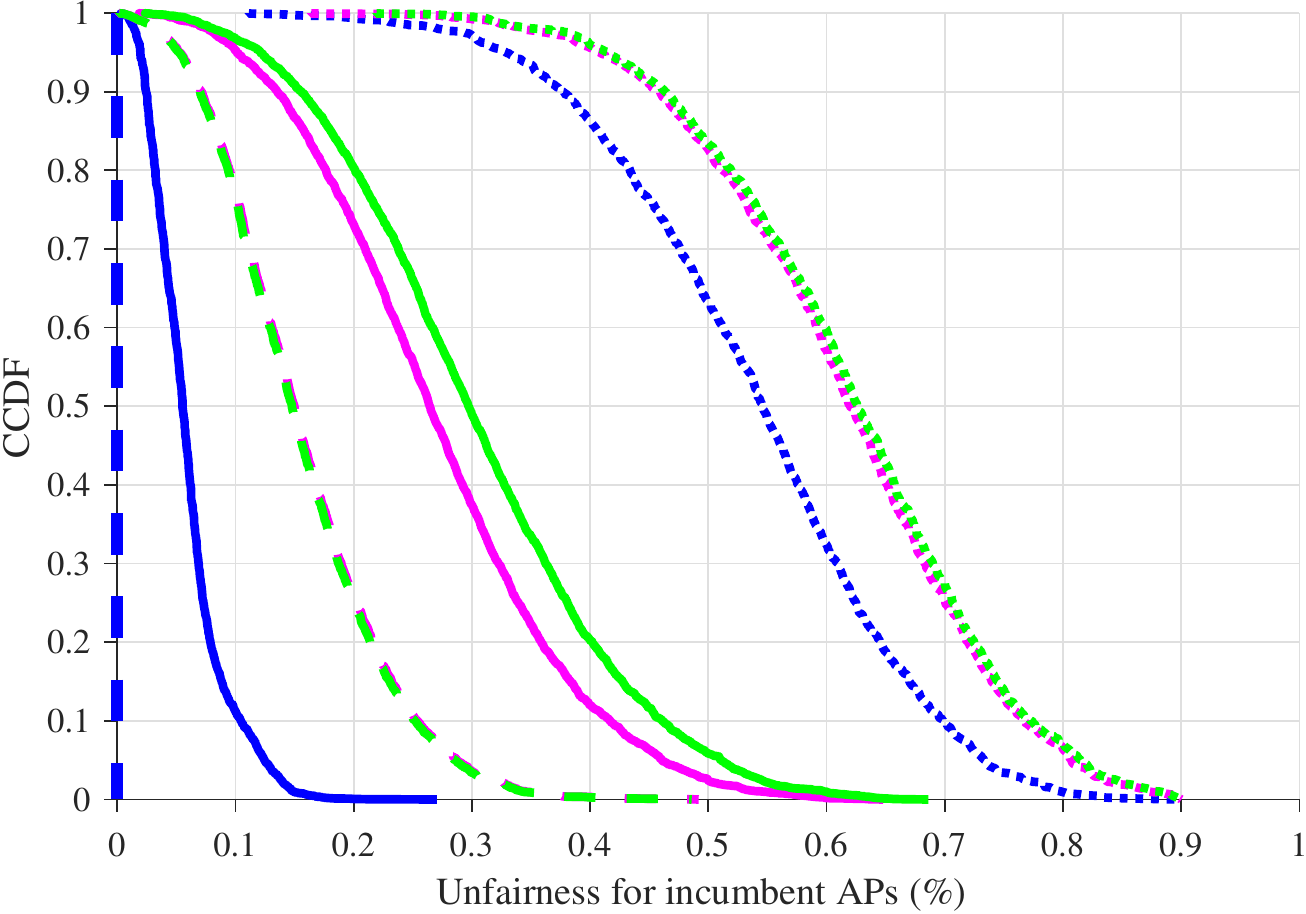} \label{fig_4b}}
\caption{Risk representation of incumbent AP performance results for the \textbf{\emph{indoor/indoor}}, \textbf{\emph{indoor/indoor without internal walls}}, and \textbf{\emph{outdoor/outdoor}} scenarios, for \emph{single channel}, for 10 incumbent and 10 entrant APs, as (a)~distribution of throughput degradation per incumbent AP with the standalone incumbents as baseline; and (b)~distribution of Jain's unfairness index for incumbent APs in each network realization.}
\label{fig_4}
\end{figure}

This section shows the benefit of risk assessment when quantifying the harm of interference in different scenarios, i.e. \emph{indoor/indoor}, \emph{indoor/indoor without internal walls}, and \emph{outdoor/outdoor}.
Fig.~\ref{fig_4} shows how different scenarios affect the incumbent throughput degradation and Jain's unfairness for \emph{single channel}, for 10 incumbents and 10 entrants. 
Importantly, Fig.~\ref{fig_4a} shows a consistent switching point between Wi-Fi and LAA or LTE-U curves across different scenarios at 40-50\% degradation. LTE-in-unlicensed is thus consistently sometimes friend and sometimes foe to Wi-Fi, regardless of the scenario.

When comparing different scenarios for a given entrant technology in Fig.~\ref{fig_4a}, we observe the following engineering insights: (i)~the lowest risk of low throughput degradation is achieved for the \emph{outdoor/outdoor} scenario and the highest risk of low degradation for the \emph{indoor/indoor} scenario \emph{without internal walls}; (ii)~the highest risk of high throughput degradation is achieved for the \emph{outdoor/outdoor} scenario and the lowest risk of high degradation for the \emph{indoor/indoor} scenario \emph{without internal walls}; and (iii)~for the \emph{indoor/indoor} scenario there is a moderate risk of high and low throughput degradation. 
This shows that the variation of incumbent throughput is highest in the \emph{outdoor/outdoor} scenario, moderate for the \emph{indoor/indoor} scenario, and low for the \emph{indoor/indoor} scenario \emph{without internal walls}. This effect is consistent with the interference conditions in each scenario. For the \emph{indoor/indoor} scenario \emph{without internal walls} where the interference is high and the APs are located close to each other, the incumbents detect more entrants and are able to better avoid strong interference by deferring to them, at the expense of sharing the channel in time. Specifically, almost all incumbents suffer a degradation of at least 20\%, and for coexistence with Wi-Fi entrants the incumbent degradation is constant and equal to 52\%, as every incumbent detects all incumbents and entrants within CS range and the MAC efficiency also changes accordingly. In the \emph{outdoor/outdoor} scenario the AP network deployment is more sparse and the users are located at a wider range of distances from the APs that they are associated with. Consequently, users close to their corresponding APs experience low risk of degradation, but users far from their corresponding APs may face hidden node problems (i.e. at least 15\% of the APs have a throughput degradation of 100\%). The interference in the \emph{indoor/indoor} scenario where the APs are separated by walls is moderate compared with the other scenarios.   

We note that in our previous work~\cite{Simic2016, Voicu2016} we observed that in the \emph{indoor/outdoor} scenario, i.e. where the incumbents are located indoors and the entrants are located outdoors, the incumbents and entrants are isolated from each other, due to the high attenuation through the external walls. The corresponding risk of interference from the entrants to the incumbents would therefore be zero, so we do not present results for this scenario in this paper. 

Fig.~\ref{fig_4b} shows Jain's throughput unfairness among incumbents for different scenarios. Consistent with our results in Fig.~\ref{fig_4a} and the corresponding discussion, the lowest unfairness is achieved for the \emph{indoor/indoor} scenario \emph{w/o internal walls} with down to zero unfairness for incumbents coexisting with Wi-Fi entrants. A moderate risk of unfairness is  shown for the \emph{indoor/indoor} scenario, whereas for the \emph{outdoor/outdoor} scenario the unfairness is large. Also, for each specific scenario, the unfairness when coexisting with \mbox{Wi-Fi} entrants is lower than when coexisting with LAA or LTE-U entrants, consistent with the values of the CS threshold that the incumbents implement.

\section{Risk Assessment from the LTE-in-unlicensed Perspective}
\label{sec_discussion}

In this section we apply risk analysis for Wi-Fi/LTE coexistence from the point of view of the LTE-in-unlicensed technology. 
In Section~\ref{results_ent} we show a selection of the throughput results for LAA or LTE-U entrants coexisting with Wi-Fi incumbents, as complementary results to those for the Wi-Fi incumbents in Section~\ref{section_results}. We note that the LTE-in-unlicensed entrant results are outside the scope of the technical policy question of whether LTE is friend or foe to Wi-Fi. However, they provide further engineering insight into Wi-Fi/LTE coexistence, which can be used by LTE-in-unlicensed operators to decide which variant to deploy for best network performance.
In Section~\ref{results_interop} we consider a further separate case of LTE-in-unlicensed inter-operator coexistence, where there are no Wi-Fi incumbents.
We thus explore the hypothetical future case where LTE is the dominant technology in the 5~GHz unlicensed band and we analyse which choice of technology is better suited from the operator perspective when coexisting with other operators.

\subsection{Wi-Fi/LTE Coexistence from the Entrant Perspective}
\label{results_ent}

In this section we present a selection of the throughput results for the LTE-in-unlicensed entrants when coexisting with Wi-Fi incumbents, for the scenarios in Section~\ref{simulation_model}. For the entrants we apply the same throughput consequence metrics as those defined for the incumbents in Section~\ref{metrics_risk}. Specifically, we consider the throughput degradation per entrant, where the baseline is the throughput of the standalone entrants (i.e. when there is no incumbent), and the unfairness among entrants.

\begin{figure}[!t]
\centering
\subfloat[]{\includegraphics[width=0.9\columnwidth]{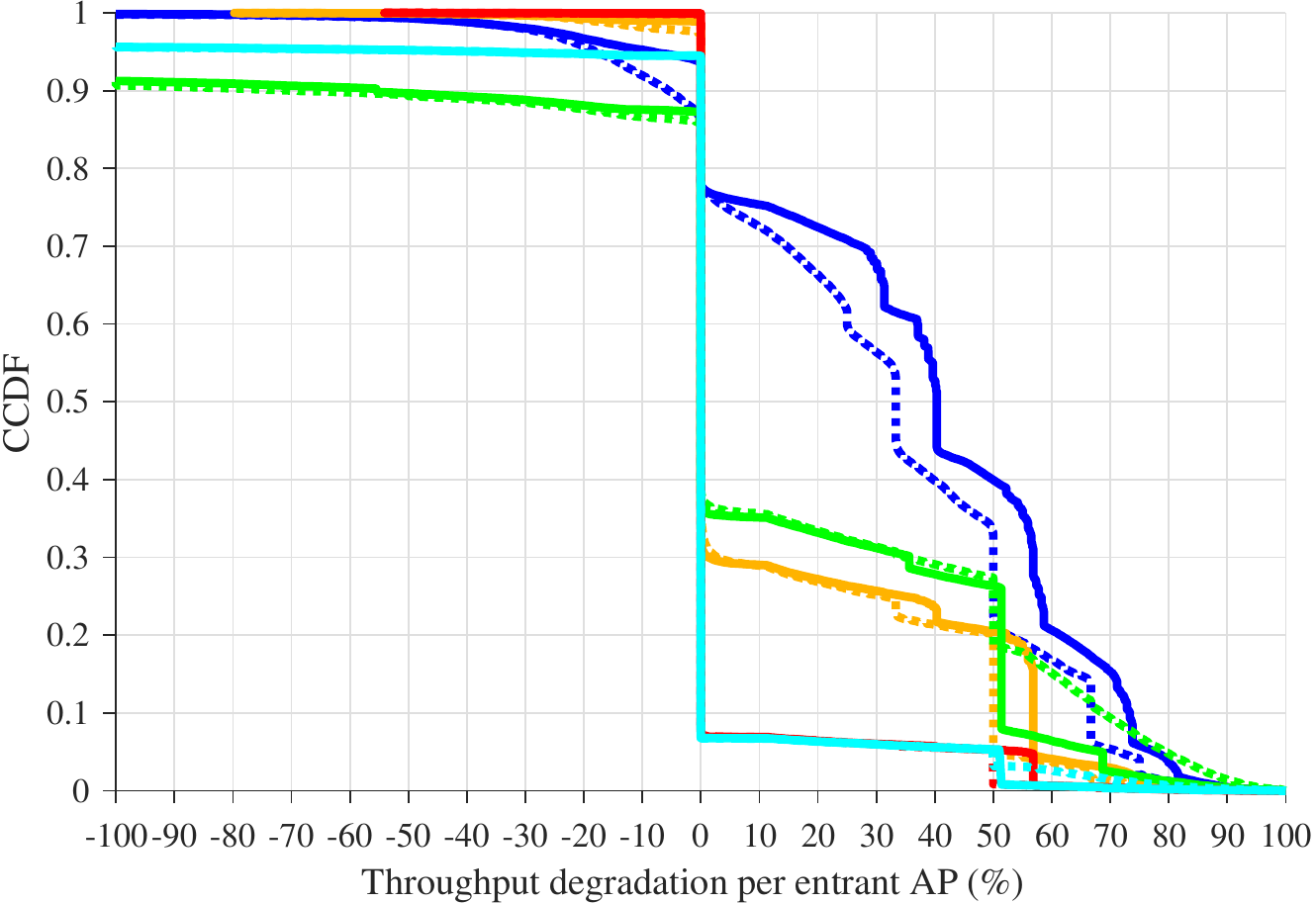} \label{fig_5a}}
\\
\subfloat[]{\includegraphics[width=0.9\columnwidth]{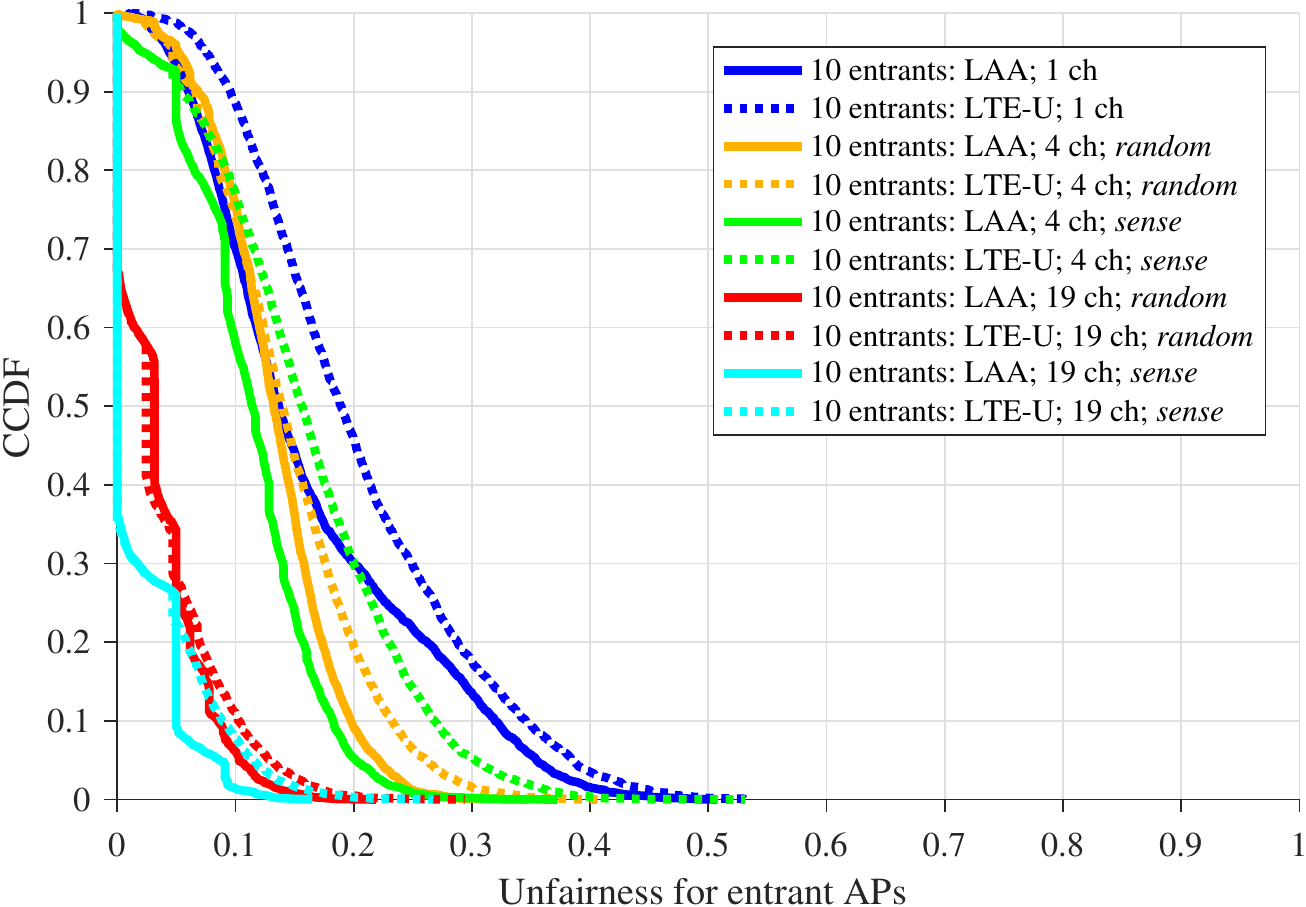} \label{fig_5b}}
\caption{Risk representation of \textbf{entrant} AP performance results for the \emph{indoor/indoor} scenario, for \textbf{different number of channels}, for 10 entrant and 10 Wi-Fi incumbent APs, as (a)~distribution of throughput degradation per entrant AP with the standalone incumbents as baseline; and (b)~distribution of Jain's unfairness index for entrant APs in each network realization.}
\label{fig_5}
\end{figure}

Fig.~\ref{fig_5} shows the risk representation  of entrant throughput results, for the \emph{indoor/indoor} scenario with 10 entrant APs coexisting with 10 incumbent APs, for different numbers of channels and channel selection schemes. 
Fig.~\ref{fig_3} shows the incumbent results for the same scenario and channel configurations.
The throughput degradation for LAA entrants in Fig.~\ref{fig_5a} is consistently but marginally higher than for LTE-U entrants, for \emph{single channel} (by up to 15 percentage points for the 70\textsuperscript{th} percentile). This shows that the LAA entrants are affected more by the coexistence with Wi-Fi incumbents, as the LAA entrants  have to defer to the neighbouring incumbents and thus transmit for a shorter time, according to the implemented LBT mechanism. Although the LTE-U entrants also reduce their duty cycle transmission time according to the number of detected incumbents, the likelihood of random overlapping transmissions from the neighbouring LTE-U entrants causing strong interference is also reduced compared to the standalone LTE-U entrants. The resulting throughput degradation is thus lower for LTE-U than for LAA.
Although this metric shows that LAA performs worse than LTE-U, we note that this is a \emph{relative} metric, which depends on the absolute throughput value for LAA and LTE-U, respectively. The entrant throughput degradation can thus be considered in practical cases for e.g. determining the risk for an LTE variant when coexisting with Wi-Fi in shared spectrum bands vs. operating in dedicated spectrum bands. For cases where the absolute throughput is more relevant, we provide further entrant results in~\cite{Voicu2016, Simic2016}.     

For a larger number of available channels (i.e. 4 or 19), LAA and LTE-U have a similar throughput degradation due to Wi-Fi incumbents, showing that both LTE-in-unlicensed variants coexist equally well with the incumbents for typical coexistence cases in the 5~GHz band.
We note that a marginal difference in throughput degradation can be observed by comparing the results for LAA or LTE-U for 4 channels \emph{random} against \emph{sense}. A fraction of 13\% of the entrants experience a negative throughput degradation (i.e. throughput increase) and 37\% a positive throughput degradation with \emph{sense}, whereas only 2\% of the entrants experience a negative throughput degradation and 30\% a positive throughput degradation with \emph{random}. The throughput degradation thus varies less with \emph{random} than with \emph{sense}.
This effect occurs due to the higher dynamics of the \emph{sense} channel selection mechanism, compared to \emph{random}.
Specifically, with \emph{random} each of the entrant APs always transmits on the same randomly selected channel, whereas with \emph{sense} each of the entrant APs selects a different channel when coexisting with incumbent APs.\footnote{We note that with \emph{sense} the entrant APs first check which channels are already occupied by Wi-Fi incumbents and then select an unoccupied channel (\emph{cf.} Section~\ref{simulation_model}), whereas for entrants implementing \emph{random}, the channel occupation is irrelevant when selecting a channel. In practice, \emph{random} and \emph{sense} would correspond to quasi-static channel allocation and frequent channel \mbox{(re-)allocation}, respectively.}
Consequently, when estimating the throughput degradation for entrants with \emph{sense}, the throughput obtained for coexistence is based on a different channel allocation than the baseline throughput for standalone entrants.
Moreover, with \emph{sense} the entrants may cause more mutual interference among themselves, by avoiding the same channels occupied by the incumbents, but the mutual interference between incumbents and entrants is thus reduced. 
As such, the overall interference at the entrants depends on the specific deployment and the channels selected by the incumbents.
For 19 available channels we observe the same trend as for 4 channels. However, the throughput degradation varies even less, due to the low number of co-channel APs.
 
The unfairness for the LTE-U entrants in Fig.~\ref{fig_5b} is marginally but consistently higher than for LAA, for \emph{single channel} (up to 0.5 difference in unfairness). This effect is caused by the random duty cycle transmissions of neighbouring LTE-U entrants, which can cause strong mutual interference, if they overlap in time. By contrast, LAA entrants implementing LBT always defer to other neighbouring transmissions and avoid such strong interference, resulting in a lower variation of the entrant throughput. A similar marginal difference between LAA and LTE-U unfairness occurs for 4 available channels. For the typical case of 19 available channels in the 5~GHz band, the unfairness is similar for LAA and LTE-U entrants, as the number of co-channel APs is considerably reduced.         

\begin{figure}[!t]
\centering
\subfloat[]{\includegraphics[width=0.9\columnwidth]{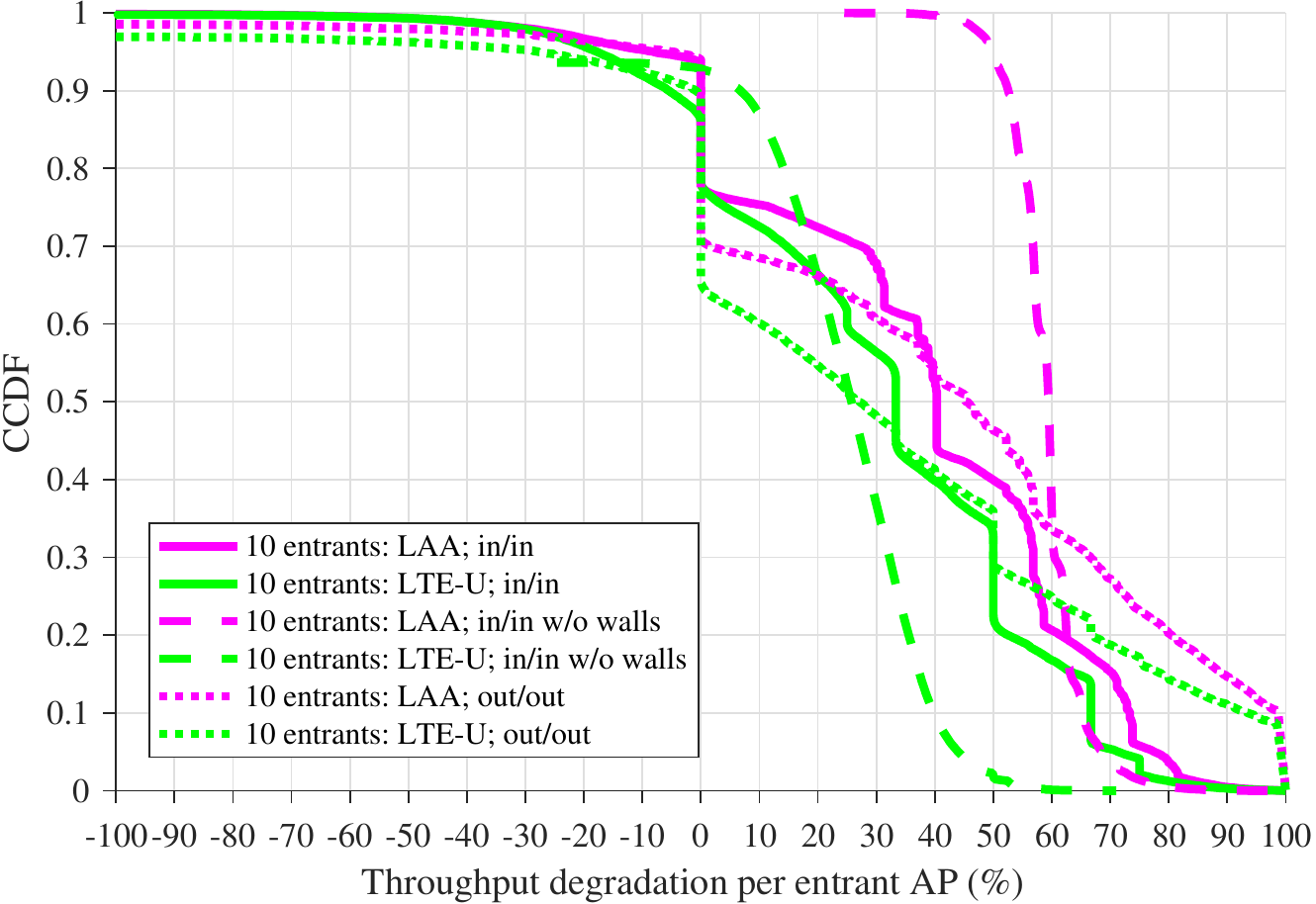} \label{fig_6a}}
\\
\subfloat[]{\includegraphics[width=0.9\columnwidth]{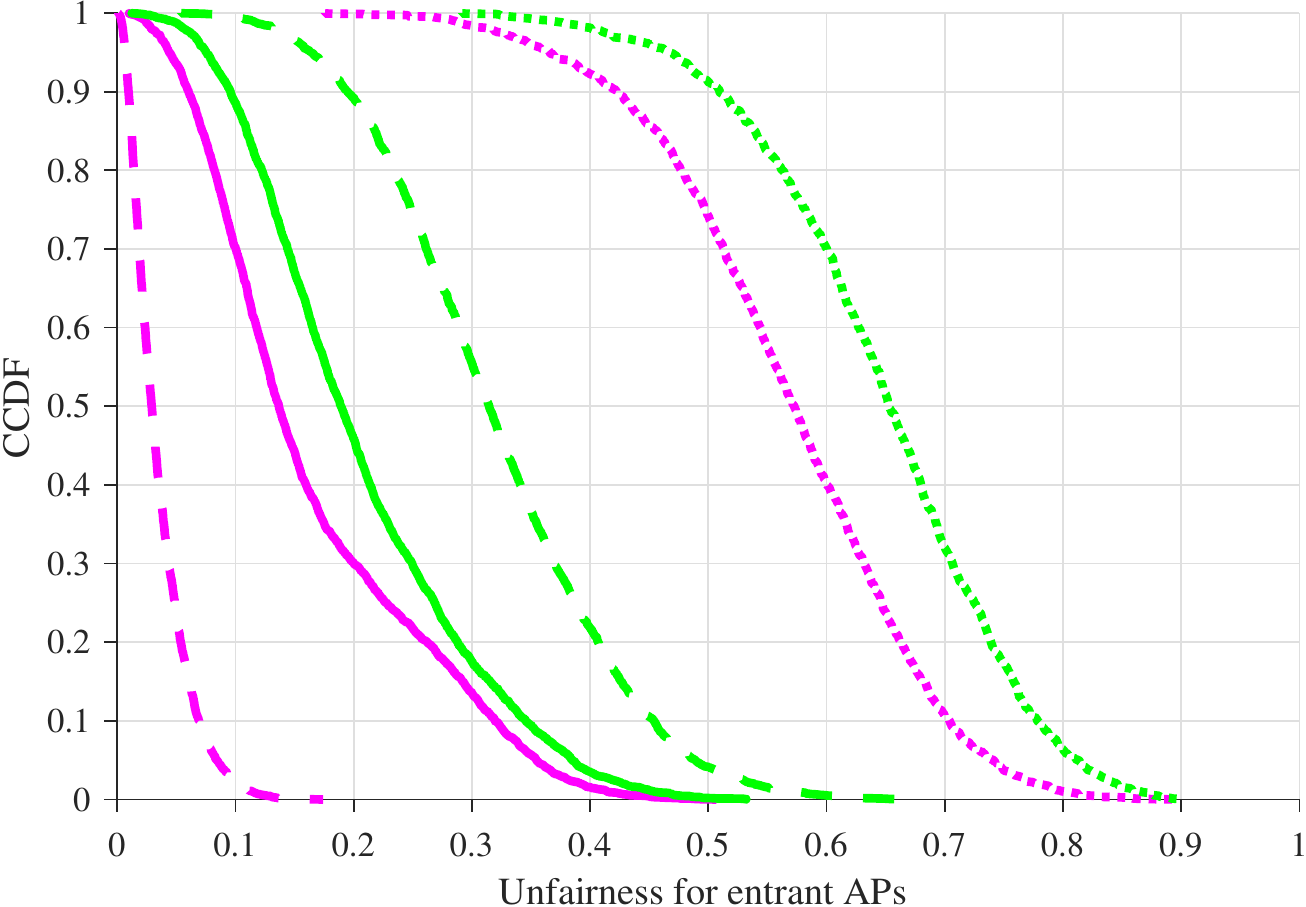} \label{fig_6b}}
\caption{Risk representation of \textbf{entrant} AP performance results for the \textbf{\emph{indoor/indoor}}, \textbf{\emph{indoor/indoor without internal walls}}, and \textbf{\emph{outdoor/outdoor}} scenarios, for \emph{single channel}, for 10 entrant and 10 Wi-Fi incumbent APs, as (a)~distribution of throughput degradation per entrant AP with the standalone entrants as baseline; and (b)~distribution of Jain's unfairness index for entrant APs in each network realization.}
\label{fig_6}
\end{figure}  

Fig.~\ref{fig_6} shows the risk representation  of entrant throughput results, for 10 entrant and 10 incumbent APs, for different scenarios and \emph{single channel} (\emph{cf.} incumbent results in Fig.~\ref{fig_4}). Fig.~\ref{fig_6a} shows that for the \emph{indoor/indoor} scenario \emph{without internal walls}, the throughput degradation for the LAA entrants is significantly higher than for the LTE-U entrants (e.g. 35 percentage points difference for the median throughput degradation). This shows that, as discussed for the \emph{indoor/indoor} scenario in Fig.~\ref{fig_5a}, the LTE-U entrants are less affected by coexistence with incumbents than LAA entrants. This difference in LAA vs. LTE-U throughput degradation is more pronounced for the \emph{indoor/indoor} scenario \emph{without internal walls} than for the \emph{indoor/indoor} scenario, due to the stronger interference among APs in the open-plan indoor scenario. The LAA entrants thus defer to more APs, whereas the LTE-U entrants already suffer from strong interference from other LTE-U entrants, such that the coexistence with Wi-Fi incumbents does not affect them to the same extent that LAA is affected. 
Moreover, we note that 6\% of the LTE-U entrants in the \emph{indoor/indoor} scenario \emph{without internal walls} have 0~Mbps throughput (i.e. the distribution curve does not reach 1 on the y axis). For the \emph{outdoor/outdoor} scenario, the LAA throughput degradation is larger than for LTE-U, consistent with the trend observed for the other scenarios. However, the difference in throughput degradation for LAA vs. LTE-U is only marginal, given the sparser deployment with fewer neighbouring APs.

Fig.~\ref{fig_6b} shows that the unfairness for the \emph{indoor/indoor} scenario \emph{without internal walls} is significantly higher for LTE-U than for LAA (i.e. unfairness of up to 0.18 for LAA and up to 0.65 for LTE-U), due to the high likelihood of strong interference among neighbouring LTE-U entrants. By contrast, the difference between the LTE-U and LAA unfairness for the \emph{outdoor/outdoor} scenario is much lower (i.e. at most 0.1 difference), consistent with the throughput degradation results in Fig.~\ref{fig_6a}. However, the unfairness for LTE-in-unlicensed entrants is overall larger for the \emph{outdoor/outdoor} scenario than the others, due to the wider range of distances between the APs and the respective associated users, which results in a larger variation of the received power. Furthermore, hidden terminal problems are more likely for this scenario.    

Overall, our LTE-in-unlicensed entrant results show that for the typically large number of available channels in the 5~GHz band, LAA and LTE-U coexist equally well with Wi-Fi incumbents. However, for dense deployments with a large number of co-channel APs, the two considered consequence metrics show different results: from the perspective of the throughput degradation, LTE-U coexists better with Wi-Fi than LAA; from the perspective of the unfairness, LTE-U coexists worse with Wi-Fi than LAA. We note, however, that these results are also due to the interactions of the entrants among themselves, according to the implemented MAC mechanism.        

\subsection{LTE-in-unlicensed Inter-Operator Coexistence}
\label{results_interop}

In this section we use risk analysis to also explore LTE-in-unlicensed inter-operator coexistence. We consider the hypothetical future case where two operators, i.e. \emph{Operator A} and \emph{Operator B}, deploy APs of the same LTE-in-unlicensed technology, i.e. either LAA or LTE-U. We thus focus on two major inter-operator coexistence cases, as determined by different regional regulatory requirements for the MAC in the 5~GHz unlicensed band. For regions where  LBT is not required, it is expected that the operators will initially deploy LTE-U, as a less complex variant of LTE-in-unlicensed, for which compliant devices are already available~\cite{FCC_Ericsson, FCC_Nokia}. For regions where LBT is required, the operators have to implement LAA, so only LAA/LAA inter-operator coexistence is possible. Importantly, such coexistence cases are of interest to operators, since inter-operator coordination cannot be achieved in a straight-forward manner and thus LTE-in-unlicensed inter-operator coexistence may pose similar problems to Wi-Fi/LTE coexistence.

We note that LAA/LTE-U inter-operator coexistence may also occur, in regions where LBT is not required. Although we do not specifically consider this case in our analysis, LAA/LTE-U coexistence is similar to the Wi-Fi/LTE-U coexistence case analysed in Sections~\ref{section_results} and \ref{results_ent}, as both LAA and Wi-Fi implement LBT with a CS threshold of -62~dBm for deferring to other LTE-in-unlicensed APs. For evaluating coexistence between other variants of LBT (e.g. with different CS thresholds) when coexisting with duty cycle devices, we encourage the reader to use our publicly-available simulation tool~\cite{iNETS2017}.    

We consider the scenarios in Section~\ref{simulation_model} with the same locations as previously. 
For the APs of Operator A we consider similar throughput consequence metrics as defined for the incumbents in Section~\ref{metrics_risk}. Specifically, we consider the throughput degradation per Operator A AP, where the baseline is the throughput of the standalone Operator A (i.e. Operator B is not active), and the unfairness among Operator A APs. We present a selection of our results for the Operator A APs, similar to the Wi-Fi incumbent results in Section~\ref{section_results} and the LTE entrant results in Section~\ref{results_ent}.

\begin{figure}[!t]
\centering
\subfloat[]{\includegraphics[width=0.9\columnwidth]{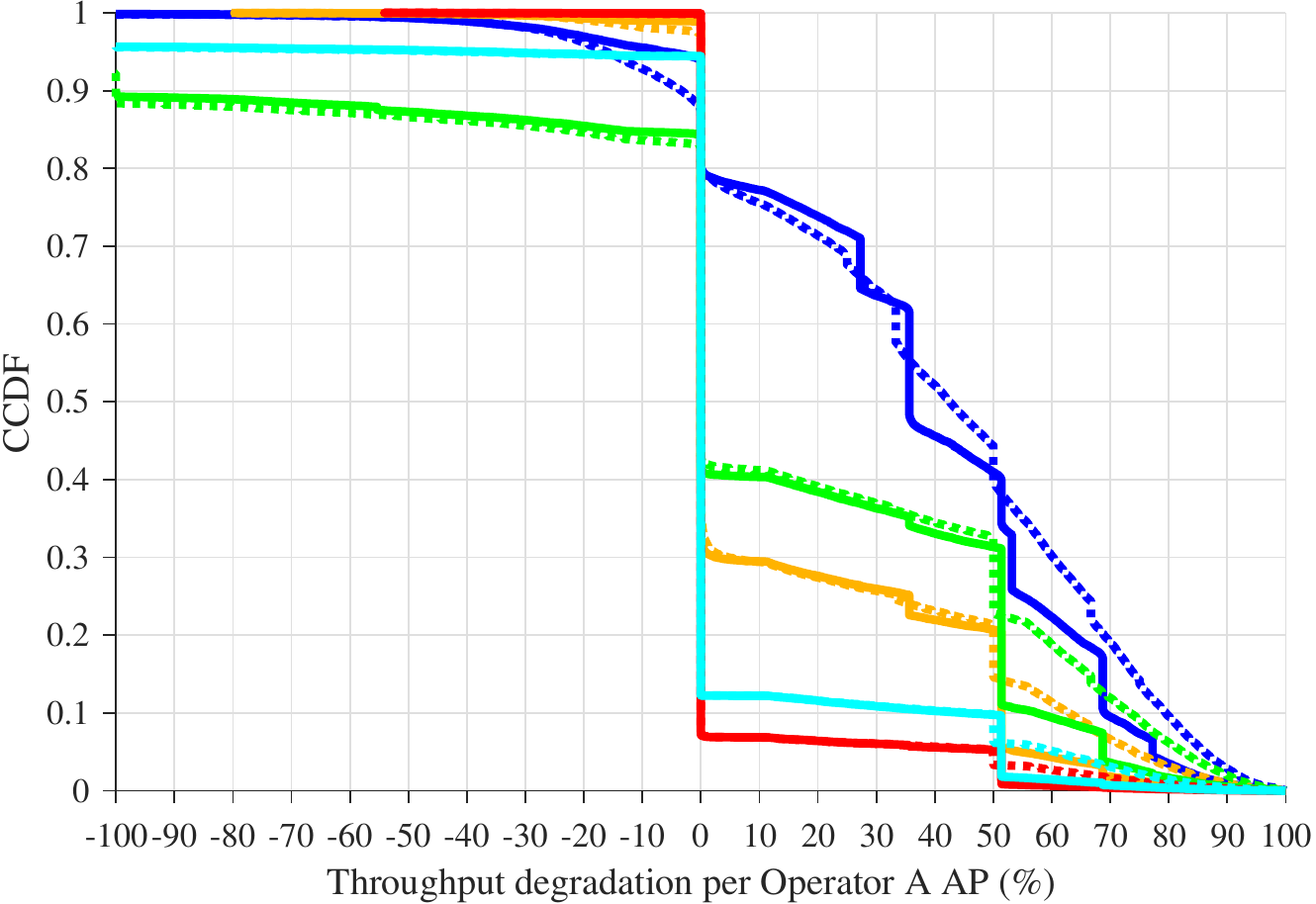} \label{fig_7a}}
\\
\subfloat[]{\includegraphics[width=0.9\columnwidth]{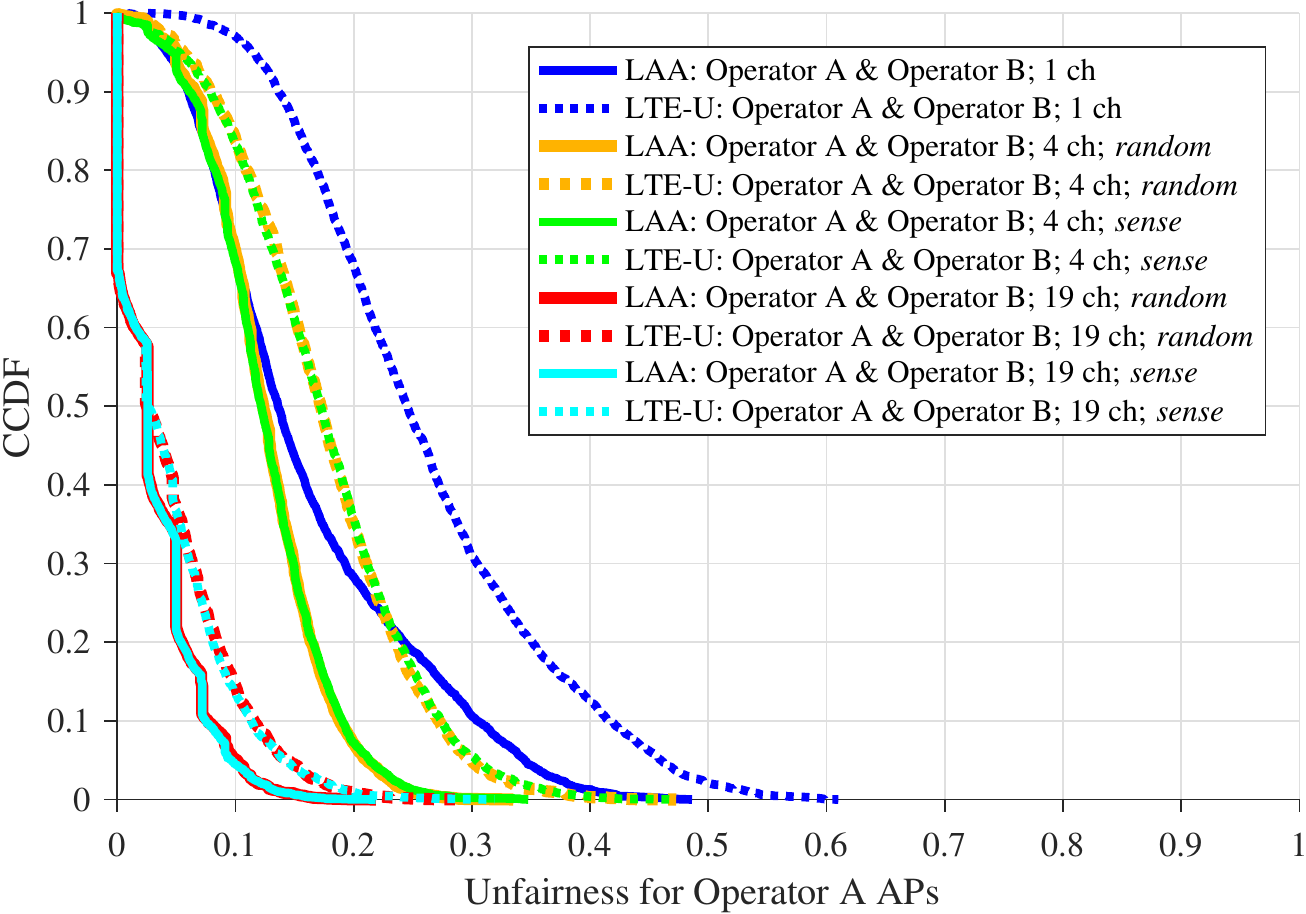} \label{fig_7b}}
\caption{Risk representation of LTE-in-unlicensed \textbf{Operator A} AP performance results for the \emph{indoor/indoor} scenario, for \textbf{different number of channels}, for 10 Operator A and 10 Operator B APs, as (a)~distribution of throughput degradation per Operator A AP; and (b)~distribution of Jain's unfairness index for the Operator A APs in each network realization.}
\label{fig_7}
\end{figure}

Fig.~\ref{fig_7} shows the risk representation  of Operator A throughput results, for the \emph{indoor/indoor} scenario with 10 Operator A APs and 10 Operator B APs, for different numbers of channels and different channel selection schemes.
Fig.~\ref{fig_7a} shows that the throughput degradation per Operator A AP is similar for LAA and LTE-U, regardless of the number of available channels. However, for \emph{single channel} and 4 available channels, the risk of high degradation (i.e. more than 50\%) is somewhat larger for LTE-U than for LAA, due to random, overlapping LTE-U duty cycle transmissions from neighbouring APs, which are avoided by LAA.
As discussed for Fig.~\ref{fig_6a}, there is a larger variation in throughput degradation for 4 channels with \emph{sense} vs. \emph{random}, due to the more dynamic \emph{sense} channel selection scheme.  

The unfairness among Operator A APs in Fig.~\ref{fig_7b} is consistently larger for LTE-U than for LAA, regardless of the number of available channels, as discussed for the unfairness in Fig.~\ref{fig_5b}. However, the difference between the LAA and LTE-U unfairness reduces when increasing the number of available channels (i.e. up to 0.12 difference for \emph{single channel} and up to 0.05 for 19 channels), due to the low number of co-channel APs.
Moreover, we note that the unfairness does not change for \emph{random} vs. \emph{sense} for LAA and LTE-U, respectively, regardless of the number of available channels. This shows that the range of long-term average throughput per AP over a given network realization is not sensitive to the dynamics of the channel selection scheme.  

In general, we observe that the choice of LBT or adaptive duty cycle MAC mechanism only marginally affects the throughput degradation of the entrants in the \emph{indoor/indoor} scenario, for the realistic cases of 4 or 19 available channels. Consequently, network operators can deploy LTE-U or LAA devices with equivalent network throughput performance. We note that the dynamics of the channel selection scheme has a slightly stronger impact on the throughput degradation than the choice of MAC mechanism. However, this impact is reflected by our throughput degradation consequence metric, due to its sensitivity to per-AP variations of the selected channel in time, whereas the unfairness capturing the long-term average throughput results over the entire network remains unchanged. 
This highlights the importance of selecting the consequence metric that reflects the engineering design goal of a particular deployment.        

\begin{figure}[!t]
\centering
\subfloat[]{\includegraphics[width=0.9\columnwidth]{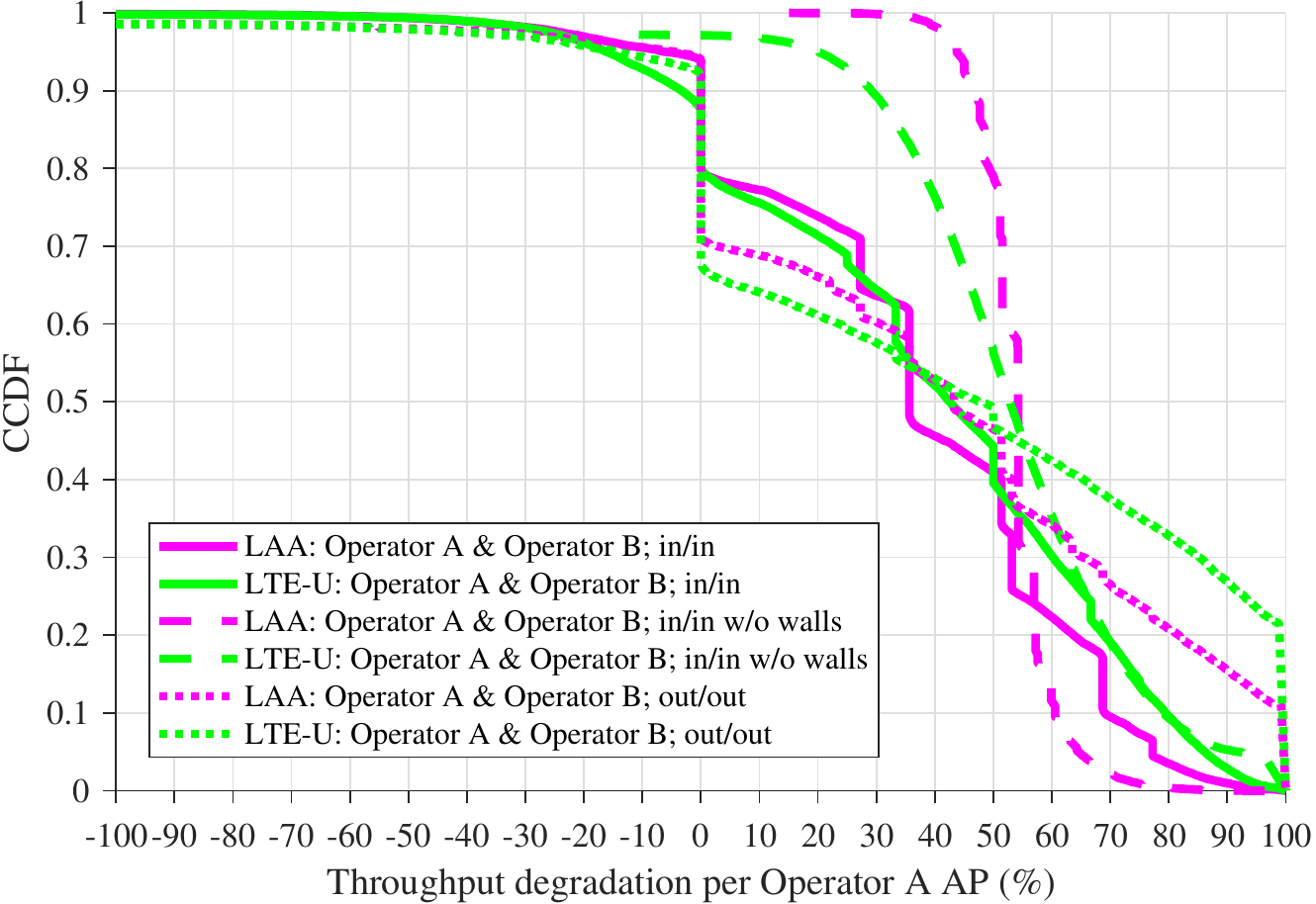} \label{fig_8a}}
\\
\subfloat[]{\includegraphics[width=0.9\columnwidth]{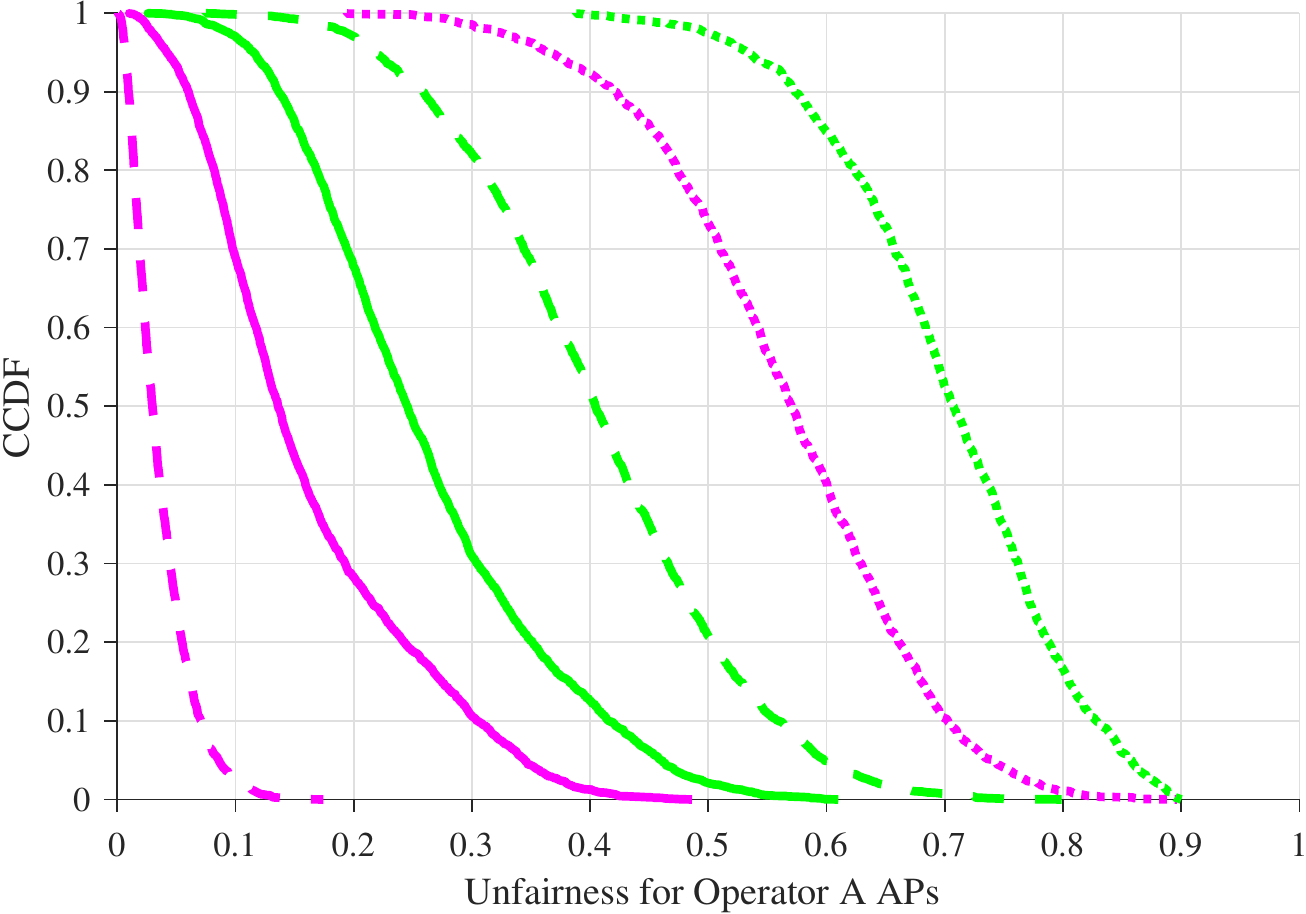} \label{fig_8b}}
\caption{Risk representation of \textbf{Operator A} AP performance results for the \textbf{\emph{indoor/indoor}}, \textbf{\emph{indoor/indoor without internal walls}}, and \textbf{\emph{outdoor/outdoor}} scenarios, for \emph{single channel}, for 10 Operator A and 10 Operator B APs, as (a)~distribution of throughput degradation per Operator A AP; and (b)~distribution of Jain's unfairness index for Operator A APs in each network realization.}
\label{fig_8}
\end{figure}  
 
Fig.~\ref{fig_8} shows the risk representation  of Operator A throughput results, for 10 Operator A and 10 Operator B APs, for different scenarios and \emph{single channel}. Fig.~\ref{fig_8a} shows that for any of the considered scenarios, there is a switching point between the throughput degradation curves for LAA and LTE-U. There are thus two regimes: of higher risk of low degradation for LAA vs. LTE-U; and of higher risk of high degradation for LTE-U vs. LAA. 
These results show that LAA implementing LBT protects the APs better against strong interference, unlike LTE-U with adaptive duty cycle, for which the APs are more likely to suffer from strong interference. 

Fig.~\ref{fig_8b} shows that the unfairness for LTE-U is significantly larger than for LAA, regardless of the scenario, consistent with the results in Fig.~\ref{fig_8a}. The largest difference in unfairness between LAA and LTE-U is observed for the \emph{indoor/indoor} scenario \emph{without internal walls} (i.e. up to 0.2 unfairness for LAA and up to 0.8 unfairness for LTE-U).        

Our results show in general that for very dense deployments with a large number of LTE-in-unlicensed APs deployed by different operators (i.e. uncoordinated co-channel APs), it is more beneficial to deploy LAA than LTE-U, regardless of the scenario, since LAA is more robust against interference and achieves a more uniform throughput among APs. 
We note that this is, of course, consistent with the use of LBT as the preferred random access MAC mechanism for high traffic load and high density scenarios (e.g. in Wi-Fi). 
By contrast, uncoordinated devices implementing adaptive duty cycle are likely to cause strong interference to each other for dense deployments and high traffic load. 
We emphasize that this result is relevant for recent regulatory discussions~\cite{U.S.FederalCommunicationsCommission2015}.
For frequency bands like the 5~GHz unlicensed band, where many channels are available, deploying LAA or LTE-U (i.e. LBT or adaptive duty cycle MAC in general) in practice results in a similar throughput performance.

\section{Conclusions}
\label{section_conclusions}
In this paper we presented a case study of Wi-Fi/LTE coexistence in the 5~GHz band, in order to demonstrate the value of risk-informed interference assessment in making regulatory decisions and for providing engineering insight. 
We applied risk assessment methods to this coexistence problem from both incumbent and entrant perspectives by (i)~identifying co- and adjacent channel interference as hazard modes, (ii)~defining the throughput degradation and Jain's throughput unfairness as consequence metrics, and (iii)~assessing the likelihood and consequence for different network densities, numbers of available channels, and scenarios (i.e. \emph{indoor/indoor}, \emph{indoor/indoor without internal walls}, and \emph{outdoor/outdoor}). We performed extensive Monte Carlo simulations for Wi-Fi incumbents coexisting with \mbox{LTE-in-unlicensed} entrants and we estimated the downlink throughput by considering co- and adjacent channel interference. 
Furthermore, we highlighted our newly publicly-available network simulation tool for risk assessment of \mbox{Wi-Fi/LTE} coexistence~\cite{iNETS2017}. 

We demonstrated that risk assessment is an effective method for evaluating the harm caused by interference in a comprehensive and intuitive manner. Our analysis clearly showed that \mbox{LTE-in-unlicensed} is neither friend nor foe to Wi-Fi in general, and thus that no regulatory intervention is needed to ensure harmonious technical coexistence. From an engineering perspective, our results showed that Wi-Fi incumbents suffer a lower risk of interference when coexisting with \mbox{Wi-Fi} entrants compared with \mbox{LTE-in-unlicensed} entrants in locally dense deployments, but 
the opposite holds for sparse deployments, due to the Wi-Fi MAC design.
Also, for the high number of available channels expected in practice, there is a negligible risk of interference for \mbox{Wi-Fi} incumbents from \mbox{LTE-in-unlicensed} entrants, which renders both policy and engineering coexistence issues largely irrelevant.
From the LTE-in-unlicensed entrant perspective, both LAA and \mbox{LTE-U} variants coexist equally well with Wi-Fi incumbents.  
For LTE intra-technology inter-operator coexistence, both variants typically coexist well in the 5 GHz band. However, for very dense deployments, LAA causes less mutual interference between operators than LTE-U, due to implementing LBT.






\bibliographystyle{IEEEtran}
\bibliography{IEEEabrv,bibliography}

\begin{thebibliography}{10}
\providecommand{\url}[1]{#1}
\csname url@samestyle\endcsname
\providecommand{\newblock}{\relax}
\providecommand{\bibinfo}[2]{#2}
\providecommand{\BIBentrySTDinterwordspacing}{\spaceskip=0pt\relax}
\providecommand{\BIBentryALTinterwordstretchfactor}{4}
\providecommand{\BIBentryALTinterwordspacing}{\spaceskip=\fontdimen2\font plus
\BIBentryALTinterwordstretchfactor\fontdimen3\font minus
  \fontdimen4\font\relax}
\providecommand{\BIBforeignlanguage}[2]{{%
\expandafter\ifx\csname l@#1\endcsname\relax
\typeout{** WARNING: IEEEtran.bst: No hyphenation pattern has been}%
\typeout{** loaded for the language `#1'. Using the pattern for}%
\typeout{** the default language instead.}%
\else
\language=\csname l@#1\endcsname
\fi
#2}}
\providecommand{\BIBdecl}{\relax}
\BIBdecl

\bibitem{VoicuBaltimore2017}
A.~M. Voicu, L.~Simi\'c, J.~P. de~Vries, M.~Petrova, and P.~M\"ah\"onen,
  ``{Analysing Wi-Fi/LTE coexistence to demonstrate de value of risk-informed
  interference assessment},'' \emph{\emph{in} Proc. IEEE DySPAN}, Baltimore,
  2017.

\bibitem{Liang2011}
Y.-C. Liang, K.-C. Chen, G.~Y. Li, and P.~M\"ah\"onen, ``{Cognitive radio
  networking and communications: An overview},'' \emph{IEEE Trans. on Vehicular
  Technology}, vol.~60, pp. 3386 -- 3407, Sept. 2011.

\bibitem{deVries2017}
\BIBentryALTinterwordspacing
J.~P. de~Vries, ``Risk-informed interference assessment: A quantitative basis
  for spectrum allocation decisions,'' \emph{Telecommunications Policy}, 2017.
  [Online]. Available: \url{http://dx.doi.org/10.1016/j.telpol.2016.12.007}
\BIBentrySTDinterwordspacing

\bibitem{iNETS2017}
\BIBentryALTinterwordspacing
\emph{iNETS inter-technology wireless coexistence risk assessment tool}.
  [Online]. Available: \url{https://www.inets.rwth-aachen.de/registration.html}
\BIBentrySTDinterwordspacing

\bibitem{Forum2015}
{LTE-U Forum}, ``{LTE-U technical report -- Coexistence study for LTE-U SDL},''
  V1.0, Feb. 2015.

\bibitem{Qualcomm2015}
\BIBentryALTinterwordspacing
Qualcomm, ``{LTE-U technology and coexistence},'' \mbox{LTE-U} Forum Workshop,
  May 2015. [Online]. Available: \url{http://www.lteuforum.org/workshop.html}
\BIBentrySTDinterwordspacing

\bibitem{3GPP2015}
3GPP, ``{Study on License-Assisted Access to unlicensed spectrum (Release
  13)},'' TR 36.889, V13.0.0, June 2015.

\bibitem{NationalCable&TelecommunicationsAssociation2015}
\BIBentryALTinterwordspacing
{National Cable \& Telecommunications Association}, ``{Comments in ET docket
  no. 15-105},'' June 2015. [Online]. Available:
  \url{https://ecfsapi.fcc.gov/file/60001078155.pdf}
\BIBentrySTDinterwordspacing

\bibitem{U.S.FederalCommunicationsCommission2015}
\BIBentryALTinterwordspacing
{U.S. Federal Communications Commission}, ``{Office of Engineering and
  Technology and Wireless Telecommunications Bureau seek information on current
  trends in LTE-U and LAA technology},'' ET Docket No.~\mbox{15-105},
  DA~\mbox{15-516}, May 2015. [Online]. Available:
  \url{https://apps.fcc.gov/edocs_public/attachmatch/DA-15-516A1.pdf}
\BIBentrySTDinterwordspacing

\bibitem{Voicu2016}
A.~M. Voicu, L.~Simi\'c, and M.~Petrova, ``{Inter-technology coexistence in a
  spectrum commons: A case study of Wi-Fi and LTE in the 5-GHz unlicensed
  band},'' \emph{IEEE J. Sel. Areas Commun.}, vol.~34, no.~11, pp. 3062--3077,
  Nov. 2016.

\bibitem{Simic2016}
L.~Simi\'c, A.~M. Voicu, P.~M\"ah\"onen, M.~Petrova, and J.~P. de~Vries, ``{LTE
  in unlicensed bands is neither friend nor foe to Wi-Fi},'' \emph{IEEE
  Access}, vol.~4, pp. 6416--6426, Sept. 2016.

\bibitem{FCCTAC2015}
\BIBentryALTinterwordspacing
{FCC Technological Advisory Council}, ``{A quick introduction to risk-informed
  interference assessment},'' V 1.00, April 2015. [Online]. Available:
  \url{http://transition.fcc.gov/bureaus/oet/tac/tacdocs/meeting4115/Intro-to-RIA-v100.pdf}
\BIBentrySTDinterwordspacing

\bibitem{deVries2015}
\BIBentryALTinterwordspacing
J.~P. de~Vries, ``{Risk-informed interference assessment: A quantitative basis
  for spectrum allocation decisions},'' \emph{\emph{in} Proc. TPRC}, Arlington,
  2015. [Online]. Available: \url{http://ssrn.com/abstract=2574459}
\BIBentrySTDinterwordspacing

\bibitem{Vries2017}
\BIBentryALTinterwordspacing
J.~P. de~Vries, U.~Livnat, and S.~Tonkin, ``{A risk-informed interference
  assessment of MetSat/LTE coexistence},'' \emph{IEEE Access}, Mar. 2017.
  [Online]. Available: \url{https://doi.org/10.1109/ACCESS.2017.2685592}
\BIBentrySTDinterwordspacing

\bibitem{Alliance2016a}
{Wi-Fi Alliance}, ``{Coexistence test plan},'' V 1.0, 2016.

\bibitem{Jain1984}
\BIBentryALTinterwordspacing
R.~K. Jain, D.-M.~W. Chiu, and W.~R. Hawe, ``{A quantitative measure of
  fairness and discrimination for resource allocation in shared computer
  system},'' \emph{\emph{submitted to} ACM Transaction on Computer Systems},
  September 1984. [Online]. Available: \url{https://arxiv.org/abs/cs/9809099}
\BIBentrySTDinterwordspacing

\bibitem{Alcatel-Lucent2009}
Alcatel-Lucent, picoChip Designs, and Vodafone, ``{Simulation assumptions and
  parameters for FDD HeNB RF requirements},'' May 2009, {3GPP TSG RAN WG4
  Meeting 51, R4-092042}.

\bibitem{Achtzehn2013}
A.~Achtzehn, L.~Simi\'c, P.~Gronerth, and P.~M\"ah\"onen, ``Survey of {IEEE}
  802.11 {Wi-Fi} deployments for deriving the spatial structure of
  opportunistic networks,'' \emph{\emph{in} Proc. IEEE PIMRC, \emph{London}},
  2013.

\bibitem{MozillaLocationService2015}
{Mozilla Location Service, August 2015. [Online]. Available:
  https://location.services.mozilla.com/downloads}.

\bibitem{IEEE2012}
\emph{{IEEE Standard for Information technology - Telecommunications and
  information exchange between systems; Local and metropolitan area networks -
  Specific requirements; Part 11: Wireless LAN Medium Access Control (MAC) and
  Physical Layer (PHY) Specifications}}, IEEE Std. 802.11, Mar. 2012.

\bibitem{3GPP2009}
3GPP, ``{E-UTRA; Radio Frequency (RF) system scenarios},'' TR 36.942 V8.2.0,
  July 2009.

\bibitem{LottRhode2001}
M.~Lott and I.~Forkel, ``A multi-wall-and-floor model for indoor radio
  propagation,'' \emph{\emph{in} Proc. IEEE VTC}, Rhode, 2001.

\bibitem{ITU-R2013}
ITU-R, ``Propagation data and prediction methods for the planning of
  short-range outdoor radiocommunication systems and radio local area networks
  in the frequency range 300 {MHz} to 100 {GHz},'' Recommendation P.1411-7,
  Sept. 2013.

\bibitem{3GPP2010}
3GPP, ``{E-UTRA; Further advancements for E-UTRA physical layer aspects
  (Release 9)},'' TR 36.814 V9.0.0, Mar. 2010.

\bibitem{Bianchi2000}
G.~Bianchi, ``{Performance analysis of the IEEE 802.11 distributed coordination
  function},'' \emph{IEEE J. Sel. Areas Commun.}, vol.~18, no.~3, pp. 535--547,
  Mar. 2000.

\bibitem{FCC_Ericsson}
\BIBentryALTinterwordspacing
\emph{Ericsson AB RBS 6402 LTE Base Station}, ``{FCC ID TA8AKRD90106083},'' FCC
  ID Database. [Online]. Available: \url{https://fccid.io/TA8AKRD90106083}
\BIBentrySTDinterwordspacing

\bibitem{FCC_Nokia}
\BIBentryALTinterwordspacing
\emph{Nokia FW2R LTE Module}, ``{FCC ID 2AD8UFW2RADPM01},'' FCC ID Database.
  [Online]. Available: \url{https://fccid.io/2AD8UFW2RADPM01}
\BIBentrySTDinterwordspacing

\end{thebibliography}
%



\end{document}